\documentclass[fleqn,usenatbib]{mnras}

\usepackage{newtxtext,newtxmath,amsmath,ulem}

\usepackage[T1]{fontenc}

\DeclareRobustCommand{\VAN}[3]{#2}
\let\VANthebibliography\thebibliography
\def\thebibliography{\DeclareRobustCommand{\VAN}[3]{##3}\VANthebibliography}

\usepackage{graphicx}	
\usepackage{amsmath}	

\title[The atmospheric chemistry of KELT-20b]{The PEPSI Exoplanet Transit Survey (PETS) IV: Assessing the atmospheric chemistry of KELT-20b}

\author[Sydney Petz et al.]{ 
Sydney Petz,$^{1}$\thanks{Email: petz.16@osu.edu}
Marshall C. Johnson,$^{1}$
Anusha Pai Asnodkar,$^{1}$
Ji Wang,$^{1}$
B. Scott Gaudi,$^{1}$ 
\newauthor
Thomas Henning,$^{2}$
Engin Keles,$^{3}$
Karan Molaverdikhani,$^{4,5,2}$ 
Katja Poppenhaeger,$^{3,6}$ 
Gaetano Scandariato,$^{7}$ 
\newauthor
Evgenya K. Shkolnik,$^{8}$ 
Daniela Sicilia,$^{7}$ 
Klaus G. Strassmeier,$^{3,6}$ %
and Fei Yan$^{9,2}$ %
\\
$^{1}$Department of Astronomy, The Ohio State University, 4055 McPherson Laboratory, 140 West 18$^{\mathrm{th}}$ Ave., Columbus, OH 43210 USA\\
$^{2}$Max-Planck-Institut f\"ur Astronomie, K\"onigstuhl 17, D-69117 Heidelberg, Germany\\
$^{3}$Leibniz-Institute for Astrophysics Potsdam (AIP), An der Sternwarte 16, D-14482 Potsdam, Germany\\
$^{4}$Universit\"ats-Sternwarte, Ludwig-Maximilians-Universit\"at M\"unchen, Scheinerstrasse 1, D-81679 M\"unchen, Germany\\
$^{5}$Exzellenzcluster Origins, Boltzmannstra{\ss}e 2, 85748 Garching, Germany\\
$^{6}$Institute of Physics \& Astronomy, University of Potsdam, Karl-Liebknecht-Str. 24/25, D-14476 Potsdam, Germany\\
$^{7}$INAF -- Osservatorio Astrofisico di Catania, via S. Sofia 78, I-95123 Catania, Italy\\
$^{8}$School of Earth and Space Exploration, Arizona State University, 660 S. Mill Ave., Tempe, Arizona 85281, USA\\
$^{9}$ Department of Astronomy, University of Science and Technology of China, Hefei 230026, China\\
}

\date{Accepted XXX. Received YYY; in original form ZZZ}

\pubyear{2023}

\begin{document}
\label{firstpage}
\pagerange{\pageref{firstpage}--\pageref{lastpage}}
\maketitle

\begin{abstract}
Most ultra hot Jupiters (UHJs) show evidence of temperature inversions, in which temperature increases with altitude over a range of pressures.
Temperature inversions can occur when there is a species that absorbs the stellar irradiation at a relatively high level of the atmospheres.
However, the species responsible for this absorption remains unidentified.
In particular, the UHJ KELT-20b is known to have a temperature inversion.
Using high resolution emission spectroscopy from LBT/PEPSI we investigate the atomic and molecular opacity sources that may cause the inversion in KELT-20b, as well as explore its atmospheric chemistry. 
We confirm the presence of Fe~\textsc{i} with a significance of 17$\sigma$. We also report a tentative $4.3\sigma$ detection of Ni~\textsc{i}. A nominally $4.5\sigma$ detection of Mg~\textsc{i} emission in the PEPSI blue arm is likely in fact due to aliasing between the Mg~\textsc{i} cross-correlation template and the Fe~\textsc{i} lines present in the spectrum.
We cannot reproduce a recent detection of Cr~\textsc{i}, while we do not have the wavelength coverage to robustly test past detections of Fe~\textsc{ii} and Si~\textsc{i}. 
Together with non-detections of molecular species like TiO, this suggests that Fe~\textsc{i} is likely to be the dominant optical opacity source in the dayside atmosphere of KELT-20b and may be responsible for the temperature inversion.
We explore ways to reconcile the differences between our results and those in literature and point to future paths to understand atmospheric variability.

\end{abstract}




\section{Introduction}
\label{section:intro}

Ultra-hot Jupiters (UHJs) are prime candidates to investigate exoplanetary atmospheric chemistry due to their short orbital periods, large radii, and high temperatures. 
These features allows us to obtain a higher frequency of observations and larger signal-to-noise ratios (SNRs) than planets that are cooler, smaller, and much farther away from their host stars.
With these high SNRs, we are able to detect trace constituents and begin to piece together a comprehensive picture of the atmospheric chemistry and dynamics of these large, hot planets.

These chemical constraints are especially important as they can provide crucial insight into giant planet formation history.
Though some of the first atmospheric characterization was performed using the \textit{Hubble Space Telescope} \citep{Charbonneau2002}, more recent literature has shown that the use of ground-based facilities can provide meaningful constraints on atomic and molecular species \citep{Line2021, Brogi2023}.
In addition to ground-based facilities, recent advancements with \textit{James Webb Space Telescope (JWST)} have revealed molecular species, notably CO$_2$ \citep{JWST} which is key for understanding the primary atmospheres of hot gas giants.

Many UHJs have atmospheric temperature inversions in which temperature increases with increasing altitude. 
This is caused by some absorber that captures the incoming stellar radiation and converts it into heat, ultimately causing a rise in temperature as the altitude increases (and pressure decreases) in the pressure-temperature (P-T) profile of the planet.
This P-T profile describes the relationship between altitude and temperature in the atmosphere of a planet and is needed to obtain precise abundances.

These temperature inversions can be determined through the presence of emission lines of atomic and molecular species through their spectra. 
According to Kirchoff's Law of thermal radiation, if this temperature inversion was not present, we would instead observe absorption lines. 
Emission lines in planetary atmospheres have been found in about a half-dozen UHJs observed to date with high-resolution emission spectroscopy 
which include WASP-33b \citep{Nugroho2017}, WASP-121b \citep{Evans2017}, WASP-189b \citep{Yan2020, Yan2022a}, MASCARA-1b \citep{Scandariato2023}, KELT-9b \citep{Kasper2021}, KELT-20b \citep{Cont2021a}, WASP-18b, and WASP-76b \citep{Yan2023}.

Though there have been claims of detected molecular species in hot planets \citep{Zilinskas2022}, there is currently no confirmed species found in the atmospheres of UHJs. Because of the high temperatures of these planets (>2000 K), molecular species are thought to be unlikely to survive in their atmospheres.
Despite this, molecules with strong optical opacity, such as TiO or VO have been invoked to be responsible for temperature inversions.
There have been claimed detections of TiO in transmission and emission in WASP-33b by \cite{Nugroho2017} and \cite{Cont2021b}, and in absorption in WASP-189b by \cite{Prinoth2022}.
However, these TiO detections in WASP-33b could not be replicated by \cite{Herman2020} or \cite{Serindag2021}. 
Additionally, VO has been detected in WASP-121b by \cite{Hoeijmakers2020} in high resolution, but their findings could not be replicated by \cite{Merritt2020} with low resolution data.
Though these findings could not be replicated, predicting its presence provided evidence for the existence of neutral vanadium.
Furthermore, VO has recently been detected independently in transmission with ESPRESSO and MAROON-X in \cite{Pelletier2023}.

Even though these species amongst others such as CaO, MgH, CaH, and AlO have been proposed to cause temperature inversions in UHJs \citep{Gandhi2019}, there is currently little direct evidence supporting strong detections of molecular inversion agents.

UHJ KELT-20b (also known as MASCARA-2b), is an exoplanet discovered independently by the Kilodegree Extremely Little Telescope (KELT) and  the Multi-site All-Sky CAmeRA (MASCARA) ground-based transit surveys \citep{Lund2017, Talens2018}. With an effective temperature of about 8720 K and a $V$-band magnitude of 7.58, its type A2 host star is one of the brightest known stars to have a transiting planet. Orbiting closely to this star with an orbital period about 3.5 days and with an effective temperature of about 2200K, KELT-20b has become a popular candidate to explore atmospheric properties.

Through the use of high-resolution emission spectroscopy, several species such as Fe~\textsc{i} \citep{Johnson2022, Yan2022b, Borsa2022, Kasper2022}, Ni~\textsc{i} \citep{Kasper2022}, Fe~\textsc{ii} and Cr~\textsc{i} \citep{Borsa2022}, and Si~\textsc{i} \citep{Cont2021a} have been detected in emission in the atmosphere of KELT-20b, indicating a strong temperature inversion. In low resolution emission, CO and H$_{2}$O have also been detected \citep{Fu2022}.
Additionally, there are multiple null detections of molecular species in KELT-20b including TiO, VO, and CaH in emission \citep{Johnson2022} and NaH, MgH, AlO, SH, CaO, VO, FeH, and TiO in transmission \citep{Nugroho2020}.
However, a low-confidence FeH detection was found in transmission in \cite{Kesseli2020}. Multiple atomic species have also been detected in transmission, including Fe~\textsc{i} \citep{Stangret2020,Gandhi2023}, Fe~\textsc{ii} \citep{Casasayas2019,Stangret2020,BelloArufe2022,Pai2022}, Ca~\textsc{ii}, Na~\textsc{i} \citep{Casasayas2019,Nugroho2020}, Mg~\textsc{i}, Cr~\textsc{i}, Mn~\textsc{i}, and Ca~\textsc{i} \citep{Gandhi2023}. \cite{Casasayas2019} also detected H~\textsc{i} Balmer line absorption.

The objectives of this paper are (1) confirm the previously reported detections of atomic and molecular species using currently the highest SNR, highest resolution emission spectroscopy dataset available, from the LBT/PEPSI, in order to maximize sensitivity;
and (2) perform a search for a broader range of atomic and molecular species.

This paper is part of the PEPSI Exoplanet Transit Survey (PETS), a large 400 hours project to use LBT/PEPSI high-resolution transmission and emission spectroscopy to study a diverse population of transiting exoplanets. Previous papers in the series set limits on the silicate atmosphere of the super-Earth 55 Cnc e \citep{Keles2022}, and studied the emission spectra of the UHJs KELT-20b \citep{Johnson2022} and MASCARA-1b \citep{Scandariato2023}. This paper is a continuation of the work started in \cite{Johnson2022}.

\section{Observations and Data Analysis}

\subsection{Observations}
The data on KELT-20b were taken with the Potsdam Echelle Polarimetric and Specrographic Instrument \citep[PEPSI;][]{Strassmeier2015} on the 2 × 8.4 m Large Binocular Telescope (LBT) located on Mt. Graham, Arizona, USA. 
We observed over an uninterrupted period on two nights of observation (May 1st, 2022 and May 18th, 2022), each lasting about 4.5 hours. 
Additionally, we observed over two bandpasses simultaneously, ranging from  4800-5441\AA\  and 6278-7419\AA\  which correspond to the blue and red arms in the spectrograph and used cross dispersers CD III and V with a resolving power of R=130,000. 
The exposure time for both observations was 300s.
All of the data collected has been previously analyzed to look for chemical inversion agents TiO, VO, FeH, and CaH.
A more detailed description of these data as well as this analysis can be found in Table~\ref{tab:Obs_table} and \cite{Johnson2022}.

\begin{table*}
	\centering
	\caption{This table displays the log of obervations taken over two periods of observations. Nspec is the number of spectra taken per period and SNR\textsubscript{red} and SNR\textsubscript{blue} are the nightly average of the 95th quantile per-pixel signal-to-noise ratios in the red and blue arms of the PEPSI spectrograph.}
	\label{tab:Obs_table}
	\begin{tabular}{lcccccr} 
		\hline
		Date (UT) & Nspec & Exp. Time (s) & Airmass Range & Phases Covered & SNR\textsubscript{blue} & SNR\textsubscript{red} \\
		\hline
		2021 May 1  & 47 & 300 & 1.01-2.03 & 0.529-0.582 & 301 & 340 \\
        2021 May 18 & 45 & 300 & 1.00-1.48 & 0.422-0.473 & 347 & 397\\

		\hline
	\end{tabular}
\end{table*}

\subsection{Systematics Correction}

Any emission signals from KELT-20b will have a magnitude that is quite small as compared to the continuum flux of the host star.
In particular, these signals may be of comparable or smaller than those due to, e.g., the telluric emission from the Earth's atmosphere and other systematic signals.
We remove the most prominent signals due to the host star, telluric lines, and general systematic noise using the methodology outlined \cite{Johnson2022}. 
See that paper for more details. 

We first remove the telluric lines present in our spectra.
We use the \texttt{molecfit} package \citep{Smette2015, Kausch2015}.
which fits a telluric spectrum to the data, and then removes the fitted telluric lines. 
This process is necessary for our data taken from the red arm of the PEPSI spectrograph, but not for the blue arm, which is largely free of tellurics.
While \texttt{molecfit} works fairly well, it doesn't work perfectly to remove telluric lines, particularly in regions with strong telluric lines.
In these regions, these strong lines are saturated, therefore making them very difficult to remove since they cover entire regions of the spectrum.
To solve this problem, we removed any parts of the spectra that have strong lines that span the red arm, causing us to fit three separate regions of this arm at 6290-6320\AA, 6470-6520\AA, and 7340-7410\AA.
This process is described in more detail in \cite{Johnson2022}.

After using \texttt{molecfit}, we take the telluric-corrected spectra, calculate a median spectrum (the spectra have already been continuum-normalized to unity by the PEPSI pipeline), and subtract the median spectrum from each time-series spectrum.
This ensures that the stellar lines and time invariant components are removed.
We then implement \texttt{Sysrem} \citep{Tamuz2005}, which removes common linear or time-varying systematics.   Our implementation of \texttt{Sysrem} is based upon \texttt{PySysRem}\footnote{\url{https://github.com/stephtdouglas/PySysRem}}, with some modifications to allow it to handle spectroscopic data.
We use \texttt{Sysrem} to remove 3 systematics in the blue arm and 1 systematic for the red arm as suggested in \cite{Johnson2022}. We use the airmass of each exposure as the initial guesses for the first systematic removed. 
By implementing \texttt{Sysrem} in addition to \texttt{molecfit}, we can remove the residual telluric absorption that \texttt{molecfit} was unable to fully fit.

\subsection{Cross Correlation Analysis}
\label{subsection:CCF}
We use a standard cross correlation methodology between the template model spectra and the cleaned observed spectra \citep{Brogi2012}, using an implementation inspired by \cite{Nugroho2017}. 
The details surrounding the creation of these model spectra is outlined in section \ref{subsec:Modelspectra} and in \cite{Johnson2022}.

We continue this process by using a grid of possible radial velocity semi-amplitude ($K_p$) values of KELT-20b and shifting the time series CCFs into the planetary rest frame assuming the given value of $K_p$.
The shifted CCFs are then stacked, combining the corrected data from both observation periods and both arms of the spectrograph.
These CCFs are combined in a weighted sum; the weights are the product of the SNR of each arm and the total equivalent width in the model emission lines within that arm. We calculate the SNR for each pixel within the spectrum, and then estimate the SNR continuum as the 95th percentile of the individual pixel SNRs; it is this latter number which is used in the CCF weights. This is, again, following \cite{Johnson2022}, and is intended to weight the CCFs by both the SNR and the information content expected for the given species.

We can then search for a peak at the expected $K_p$ and $v_{sys}$ parameters and evaluate its SNR using the resulting CCF map. 
We determine the significance of this value by estimating the standard deviation of the points given in the CCF map at $|v_{sys}|$ >100 km/s. 
This significance is valid if the noise across the CCF map is consistent across all $v$ values - which is not guaranteed - and if the noise follows a Gaussian distribution.
To account for this noise, we adopt a 4$\sigma$ threshold for a detection in order to be conservative. 
However, if our data peak is between 4-5$\sigma$, we label it as a "tentative detection", and delve into it further to test its reliability. 
To explore these detections, we examine the data peak in both arms and both observation periods to determine whether the data peak appears consistently, and within the right planetary rest frame - as a real detection should.

To validate our results, we can perform injection-recovery tests in which we inject the generated model for a given species into the collected data, and then attempt to recover it using the procedures outlined above. 
This is done after the spectra have been corrected by \texttt{molecfit}, but before they have been run through \texttt{Sysrem}, allowing us to treat the spectrum and injected model exactly the same way we treated our original data.
The significance of these tests are computed using the same procedures as with our original data. We expect the significance of these tests to be much higher than with our data as the generated models exactly match the template model spectra. 
Since \texttt{molecfit} is very computationally complicated and already successfully removes most of tellurics, we don't expect to find different results in our injection-recovery tests by also running \texttt{Sysrem}.
We can use these injection-recovery tests to help determine if our non-detections occur because our data quality is not sufficient to allow for a detection, or if it is because the species is not present in the atmosphere at the expected concentration.

\subsection{Systemic Radial Velocity}

In order to assure accurate results from our cross-correlation analysis, it is also necessary to measure the stellar radial velocity and ensure that all of the analyzed spectra are in the stellar rest frame.
We used all of our available PEPSI data for this analysis, namely the two emission spectroscopy datasets described earlier plus a transmission spectroscopy dataset detailed in \cite{Johnson2022}. This dataset consisted of 23 spectra obtained during and adjacent to transit on 2019 May 4 UT.

We used the same methodology as in \cite{Pai2022b}. Briefly, we extract the stellar line profile from each of the time-series spectra using least squares deconvolution, as described in \cite{Kochukhov2010} and \cite{Wang2017}. This used a stellar model spectrum generated using \texttt{Spectroscopy Made Easy} \citep{ValentiPiskunov1996,ValentiPiskunov2012} from a stellar line list from the Vienna Atomic Line Database (VALD3)\footnote{\url{http://vald.astro.uu.se/~vald/php/vald.php}}. We fit an analytic rotationally-broadened line profile calculated as described in \cite{Gray} to the extracted line profiles in order to measure the stellar radial velocity (RV) from each spectrum. 
We fit a Keplerian model to the RVs in order to account for the reflex motion of the star due to the orbiting planet, using a Markov chain Monte Carlo (MCMC) framework. We assume a circular orbit, and the overall RV offset is the mean stellar velocity that we need for the atmospheric analysis.
We measure a stellar RV semi-amplitude of $K_{\star}=-0.08_{-0.23}^{+0.22}$ km s$^{-1}$, consistent with zero, and with the \cite{Lund2017} $3\sigma$ upper limit of $K_{\star}<0.31$ km s$^{-1}$. Despite our much higher SNR and resolution than the spectra used in \cite{Lund2017}, we have only limited phase coverage with no spectra near quadratures, resulting in a comparable limit.
We also compute a mean stellar velocity of $v_{\star}=-22.78 \pm 0.11$ km s$^{-1}$ in the PEPSI frame. We Doppler shift our spectra accordingly and conduct all further analyses in the stellar rest frame; therefore, in our further analysis, the planetary signal should appear at $v_{\mathrm{sys}}=0$ (modulo dynamical effects from the atmosphere).

\section{Assessing Detectability}
\label{sec:Analysis} 
\subsection{Model Spectra} \label{subsec:Modelspectra}
\begin{figure}
	\label{fig:ptprofile}
    \includegraphics[width=\columnwidth]{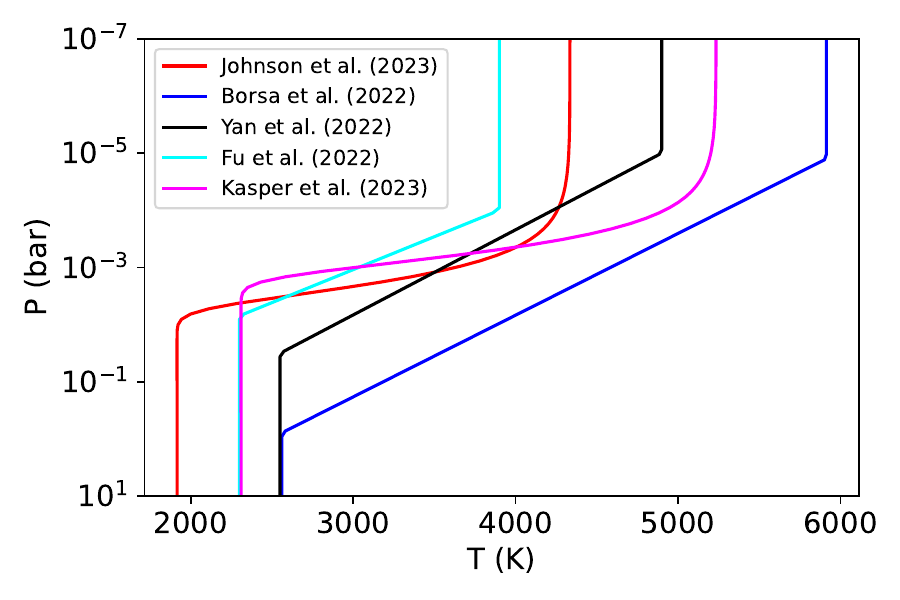}
    \caption{Pressure-temperature profiles for KELT-20~b used by previous works on the planet. We adopt the P-T profile used by \protect\cite{Johnson2022} (red) for most of the analysis in this paper.}
\end{figure}

In order to determine whether or not a certain species is present in the atmosphere of KELT-20b, we must create sets of model spectra as cross-correlation templates to compare with the spectra collected by PEPSI/LBT.
This was done using the \texttt{petitRADTRANS} package \citep{Molliere2019}, assuming stellar and planetary parameters from \cite{Lund2017} and the same P-T profile recovered for the dayside atmosphere of KELT-20b used by \cite{Johnson2022}. 
This is a model with a Guillot profile
\citep{Guillot} with $\gamma = 30 $, $\kappa$\textsubscript{IR} = 0.04 where $\gamma$ represents the ratio
between the opacity in the infrared and the optical and $\kappa$\textsubscript{IR} represents
the infrared opacity \citep{Johnson2022}. We show this profile, along with others from the literature for KELT-20~b, in Fig.~\ref{fig:ptprofile}.

To create model spectra, we chose to infer a volume mixing ratio (VMR) for tested species using the \texttt{FastChem} equilibrium chemical model \citep{FastChem}, assuming a solar abundance mixture and the same assumed Guillot P-T profile. 
We assume our VMR to be constant as a function of altitude, and take the VMR of each species when the pressure is 1 bar.
For atomic species not included in \texttt{FastChem}, we assume a solar abundance and that all of the species are in their atomic form; we thus assume the maximum possible abundance. Using solar chemical composition values from \cite{Gray} we can solve for the VMR using

\begin{equation}
\label{eqn:VMR}
    VMR_X = A_X/(A_{\mathrm{H}} + A_{\mathrm{He}} + A_X)
\end{equation}

 in which A represents the absolute abundance of the species. We can neglect the small contributions elements other than H and He to the total number density and return an estimate for the fraction of atoms of that element relative to H.

\begin{table*}
	\centering
	\caption{This table displays each notable tested species (as noted by their previously published detections and/or high detectability values) with their corresponding volume mixing ratio computed using \texttt{FastChem}, detectability values (as defined in the text), SNR given by CCFs, recovered SNR from injection-recovery tests, the data type that was used to detect the species in previous literature (transmission or emission indicated with a T or E), and the previous literature that has detected this species. The previous literature is referenced as Ref 1: \protect\cite{Stangret2020}, 2: \protect\cite{Johnson2022}, 3: \protect\cite{Yan2022b}, 4: \protect\cite{Borsa2022}, 5: \protect\cite{Casasayas2019}, 6: \protect\cite{Nugroho2020}, 7: \protect\cite{BelloArufe2022}, 8: \protect\cite{Cont2021a}, 9: \protect\cite{Kasper2022}, 10: \protect\cite{Gandhi2023}.}
	\label{tab:testedspecies}
	\begin{tabular}{lcccccccr} 
		\hline
		Species & VMR & Log Detectability Value & SNR ($\sigma$) & Injection Recovery Returned SNR ($\sigma$) & Data Type & Reference \\
		\hline
		Fe~\textsc{i}  & $5.4\times10^{-5}$ & 0.76 & 17.0 & 76.3 & T and E & 1, 2, 3, 4, and 10 \\
        Fe~\textsc{ii} & $2.7\times10^{-15}$ &  - & 0.2 & 0.2 & T and E & 4, 5, 6 and 7 \\
        Cr~\textsc{i} & $7.4\times10^{-7}$ & -0.14 & 2.2 & 27.4 & T and E & 4 and 10 \\
        Si~\textsc{i} & $1.8\times10^{-9}$ & - & 0.2 & 0.2 & E & 8  \\
        Ni~\textsc{i} & $2.8\times10^{-6}$ & -0.24 & 4.3 & 20.4 & T and E & 9 and 10 \\
        Mg~\textsc{i} & $6.8\times10^{-5}$ & -0.55 & 4.5 & 11.23 & T & 10 \\
        VO & $7.9\times10^{-9}$ & 0.001 & 3.4 & 11.5 & - & - \\
        NaH & $7.0\times10^{-9}$ & -1.64 & 2.1 & 7.8 & - & - \\
        CaH & $3.4\times10^{-7}$ & -0.48 & 2.4 & 27.4 & - & - \\
        MgH & $1.1\times10^{-8}$ & 0.72 & 1.8 & 29.9 & - & -\\
        TiO & $1.4\times10^{-7}$ & 1.72 & 2.6 & 131.8 & - & - \\
		\hline
	\end{tabular}
\end{table*}

Another necessary component to the creation of these model spectra is the opacity data for each species. The \texttt{petitRADTRANS} package contains these data for a variety of neutral and ionized atoms and molecules \footnote{\url{https://keeper.mpdl.mpg.de/d/e627411309ba4597a343/}}, but not every species of interest.
In order to either confirm or refute detections made in previous literature (particularly Ni~\textsc{i}), it was necessary to take data from the DACE Opacity Database \footnote{\url{https://dace.unige.ch/opacityDatabase/}} and convert it into a format usable for \texttt{petitRADTRANS}. This conversion was performed using a set of Fortran scripts (P. Molliere, private communication).
We chose what atomic species to consider based on the the species that were not available in \texttt{petitRADTRANS}, but were tested for detections in \cite{Kesseli2022} with the Echelle
Spectrograph for Rocky Exoplanets and Stable Spectroscopic Observations (ESPRESSO).
Since the PEPSI bandpass is a subset of the ESPRESSO bandpass, we chose not to consider species that were not searched for in \cite{Kesseli2022} as they do not have lines in the optical and/or have extremely low VMRs.
Our tested molecular species (MgH and NaH) were considered as they were predicted to be potential thermal inversion agents in \cite{Gandhi2019}.

Each set of atomic opacity data used from DACE was originally data taken from the Kurucz line list \footnote{\url{http://kurucz.harvard.edu/}}, chosen specifically as it comes from the same source as several \texttt{petitRADTRANS} atomic opacities, and to remain consistent with past results \citep{Johnson2022}. 
These data span temperature values ranging from 2500K-6100K, which includes most of the range spanned by UHJ atmospheres.
Unlike \texttt{petitRADTRANS}, these opacities use only one value for pressure - $1\times10^{-8}$ bars. 

Due to this, the pressure broadening profiles will not be exactly correct for these species.
To test whether or not this may have affected our results, we repeated our entire methodology comparing our results for Fe~\textsc{i} using opacity data from \texttt{petitRADTRANS} and opacity data from DACE. 
The resulting CCF maps had no noticeable differences, therefore we do not expect a difference in signals from the two opacity sources.

We used two sets of molecular opacity data from DACE to assess the detectability of MgH and NaH.
The MgH data were taken from the Yadin line list \citep{Yadin2012} due to its reduced file size, and the NaH data were acquired from the Rivlin line list \citep{Rivlin2015} as it was the only available data source.
Both sets of data span 50K-2900K, which includes the proposed equilibrium temperature for KELT-20b given by \cite{Lund2017} and span pressure values ranging from $1\times10^{-8}$-$1\times10^{-3}$ bars.
Because DACE provides a range of pressure values for molecular species, we expect more accurate pressure broadening profiles than given by atomic sources.

By implementing data from DACE, we were able to extend the amount of species tested from 28 to 67, giving us the ability to not only test notable species, but also investigate the detectability of a range of atomic and molecular species using generated atmospheric models.

\subsection{Quantitative Assessment of Detectability}
\label{sec:quant}

Inspired by \cite{Kesseli2022}, we assessed the detectability of these tested species quantitatively.
While we used a very similar procedure as was used in \cite{Kesseli2022}, our analysis was tailored to the PEPSI spectrograph instead of the ESPRESSO spectrograph on the Very Large
Telescope (VLT).
Because PEPSI spans 4800-5441\AA\  and 6278-7419\AA\ and ESPRESSO covers 3782-7887\AA\ , species observability may be different between the two instruments.
Additionally, our procedure differs from \cite{Kesseli2022}, which considers atomic species, but not molecular species.
To assign values to each species, it was necessary to evaluate the strength of the model emission spectra generated by \texttt{petitRADTRANS}.

\begin{figure*}
	\label{fig:periodictable}
    \includegraphics[width=\columnwidth]{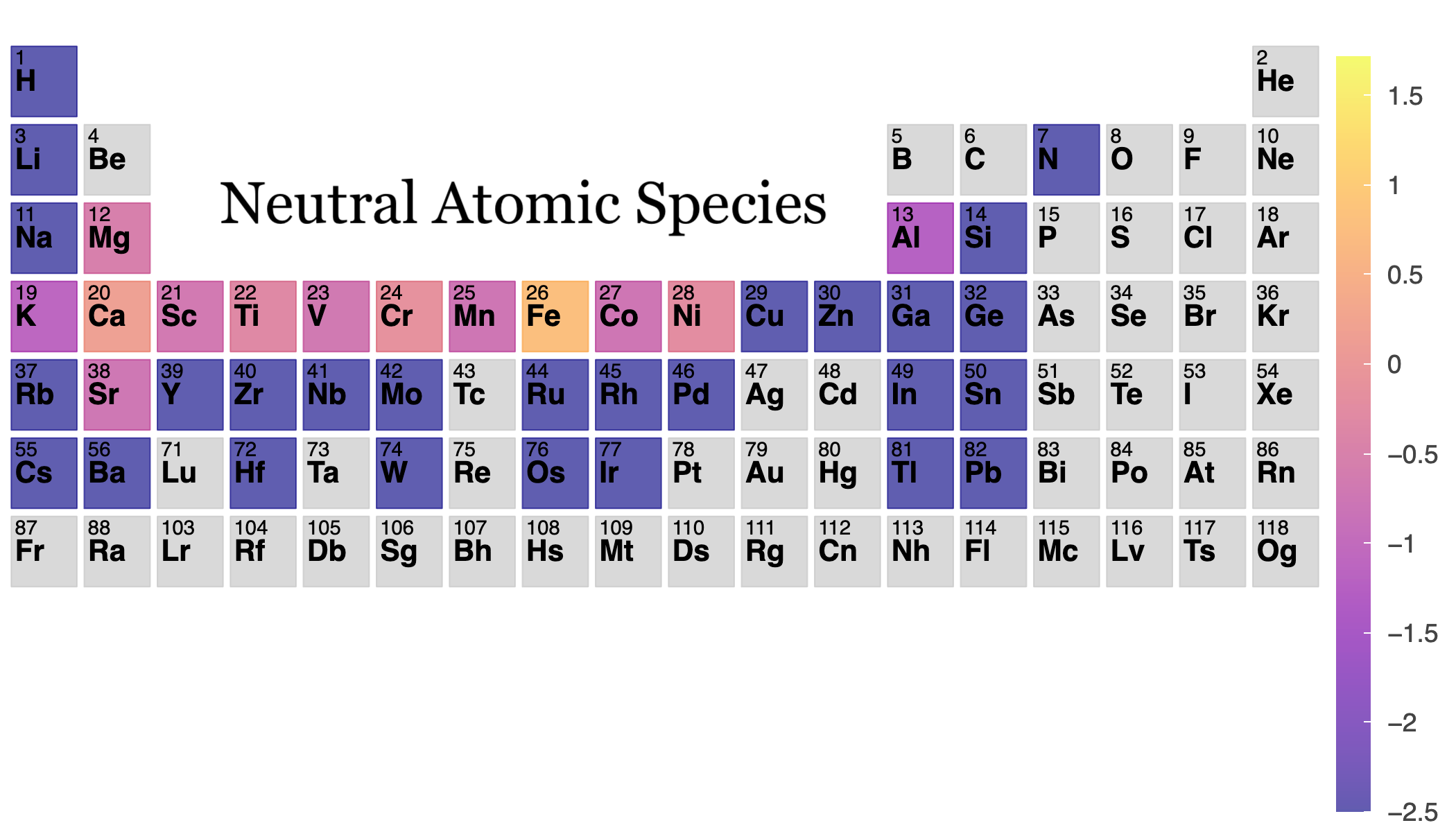}\includegraphics[width=\columnwidth]{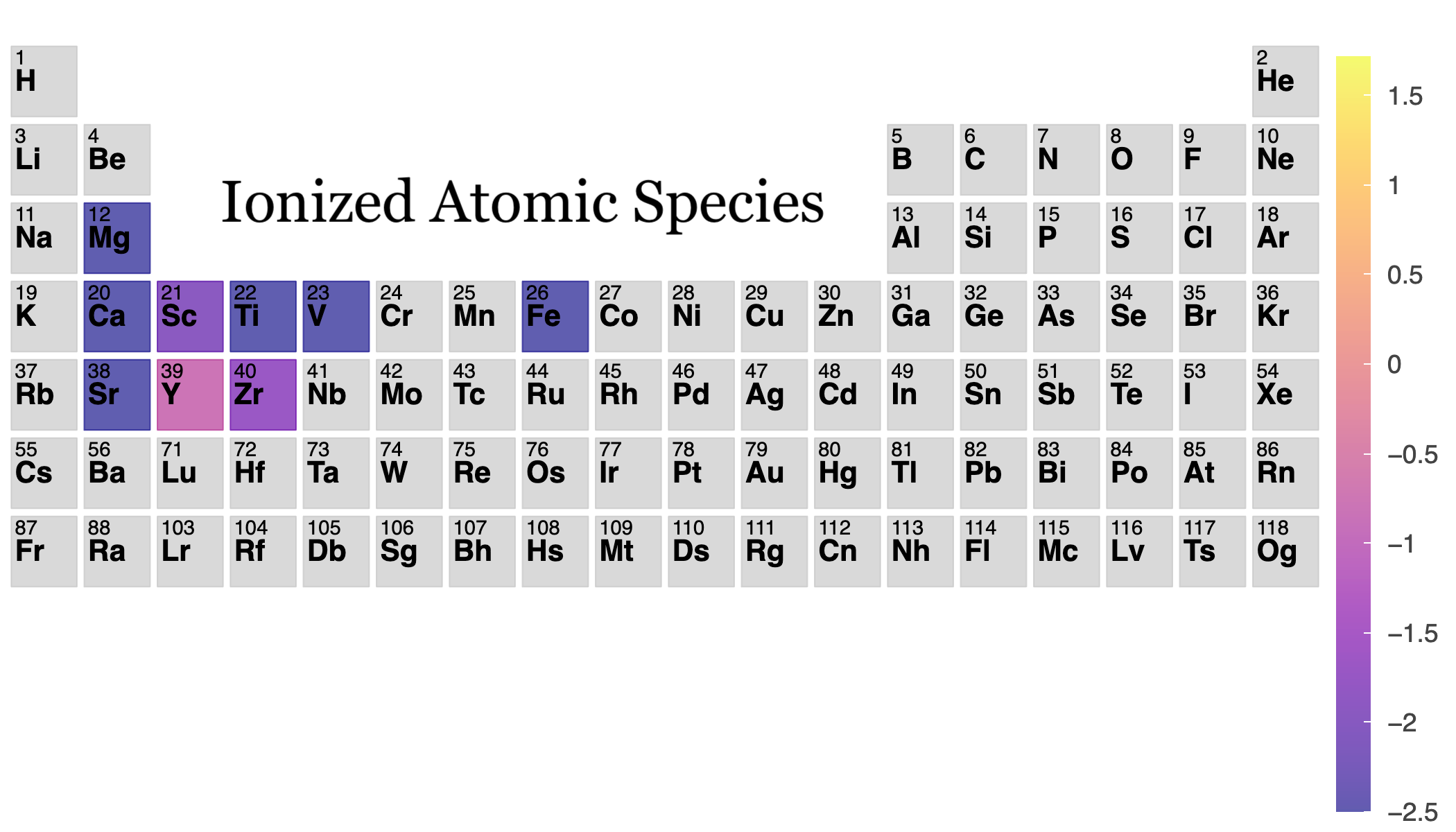} \\
    \includegraphics[width=\columnwidth]{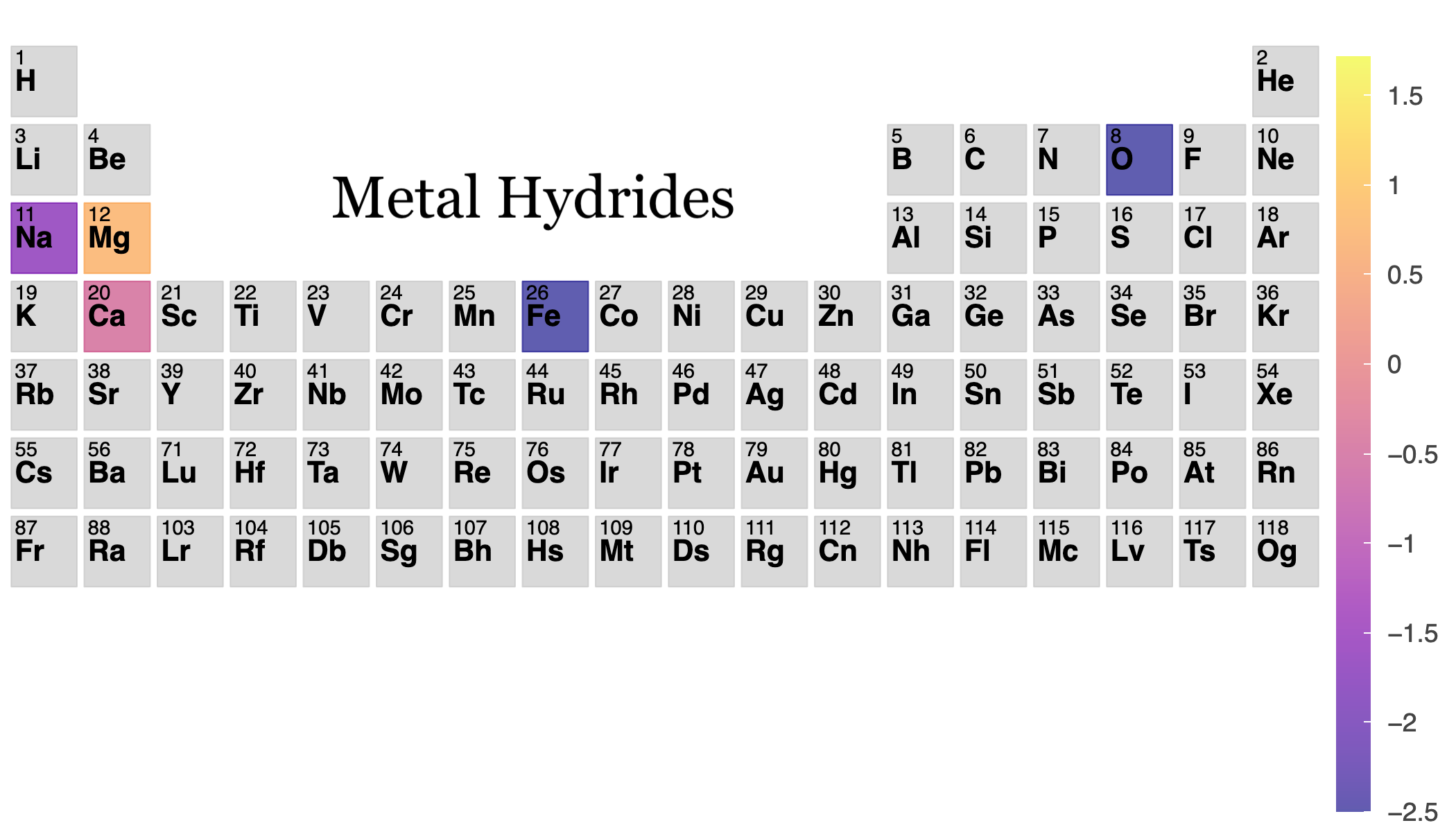}\includegraphics[width=\columnwidth]{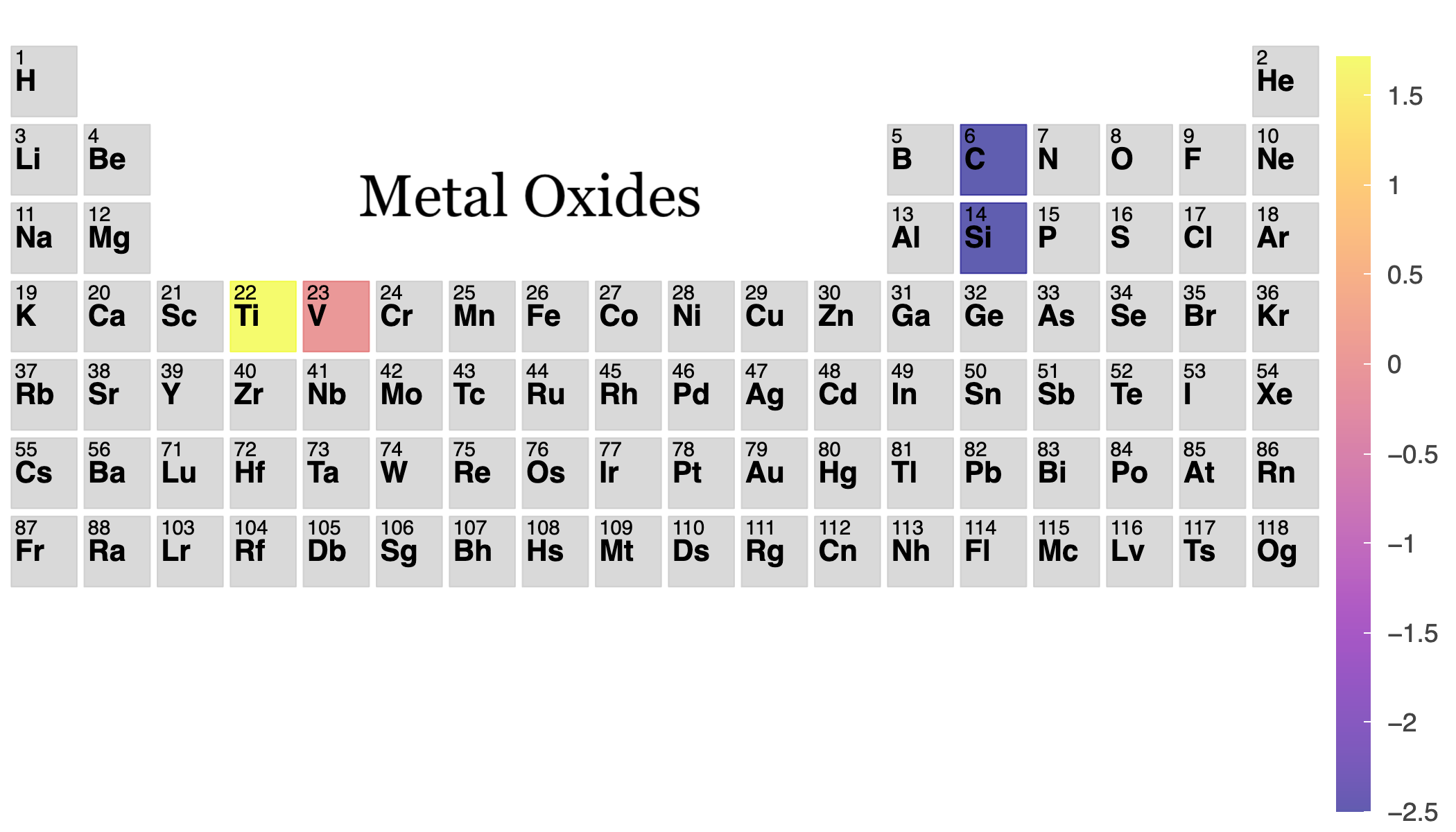}
    \caption{Periodic table plots displaying the log detectability of neutral atomic species, singly ionized atomic species, metal hydrides, and metal oxides within the atmosphere of KELT-20b. Detectability values are assessed using the methodology in the main text. Gray species indicate that they have not been considered (as to remain consistent with \protect\cite{Kesseli2022})}

    \label{PeriodicTable}
\end{figure*}

To do this, we calculated the total sum of the values of the model spectra after continuum removal, which is essentially the total equivalent width of the lines in the spectra, albeit with arbitrary units.
This returned value is indicative of how likely we are to detect each species in the atmosphere, taking into account both the abundance of the species and its line strength. These values span many orders of magnitude, so we work with the (base-10) logarithm of these values for better qualitative comparison.
We can then take the species that are more likely to produce detectable signals and proceed with a CCF analysis. 
These detectability values can be displayed in periodic table plots generated with the \texttt{ptable\_trends} code\footnote{\url{https://github.com/arosen93/ptable_trends}} as shown in Figure~\ref{PeriodicTable}.

\begin{figure*}
	\includegraphics[width=0.5\textwidth]{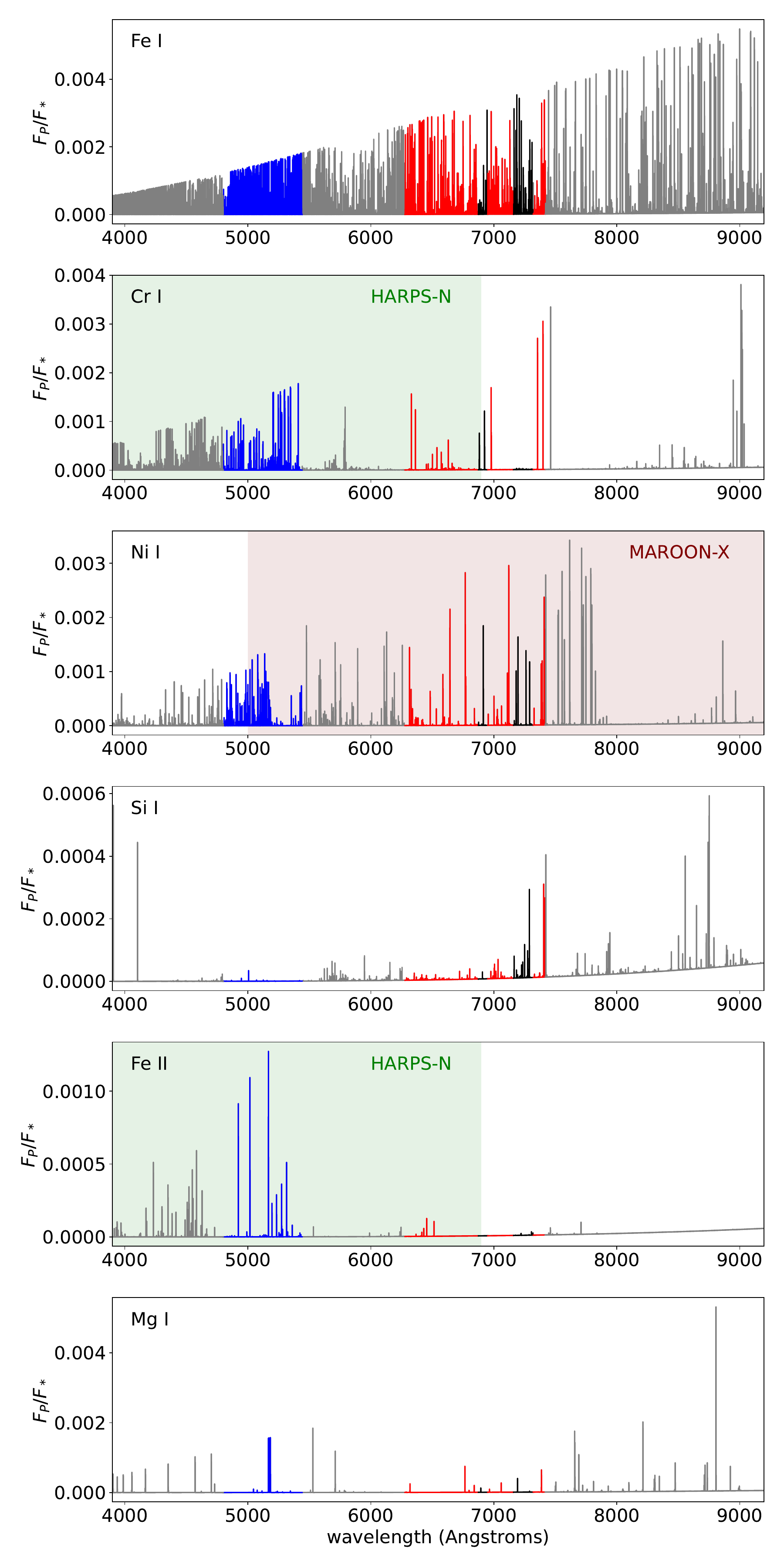}\includegraphics[width=0.5\textwidth]{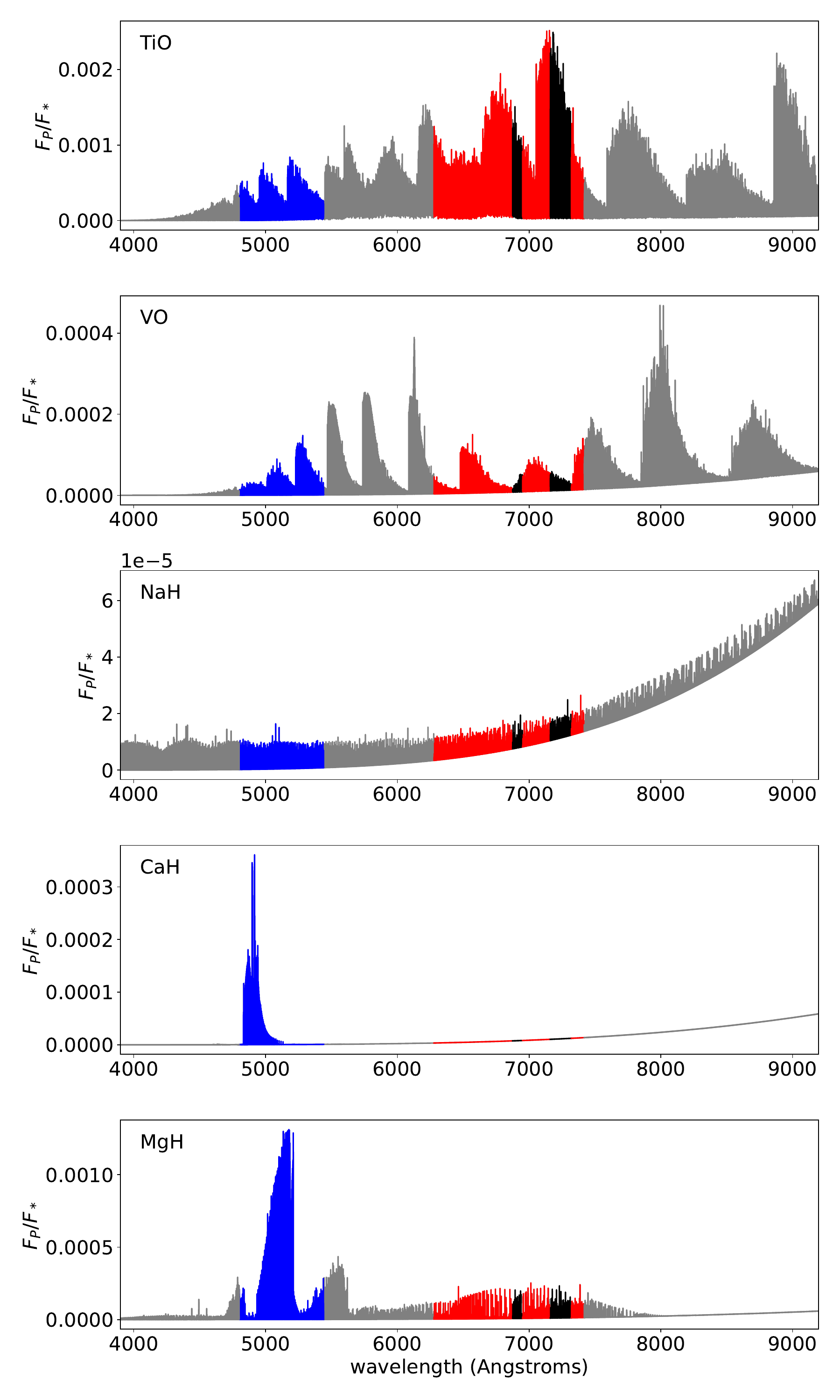}
    \caption{Model spectra of notable atomic and molecular species generated by \texttt{petitRADTRANS}. The two PEPSI bandpasses are shown in blue and red sections that correspond to their arm in the spectrograph. The HARPS-N and MAROON-X bandpass ranges are shaded as green and maroon respectively for the species previously detected with those instruments \protect\citep{Borsa2022,Kasper2022}. The CARMENES infrared bandpass used by \protect\cite{Cont2021a} to detect Si~\textsc{i} is off the right edge of the plot and is not depicted. Black sections indicate regions where parts of our data were removed due to saturated telluric lines.}

    \label{ModelSpectra}
\end{figure*}

\section{Results}

\subsection{Comparing to previous works}

\begin{figure*}
    \includegraphics[width=\columnwidth]{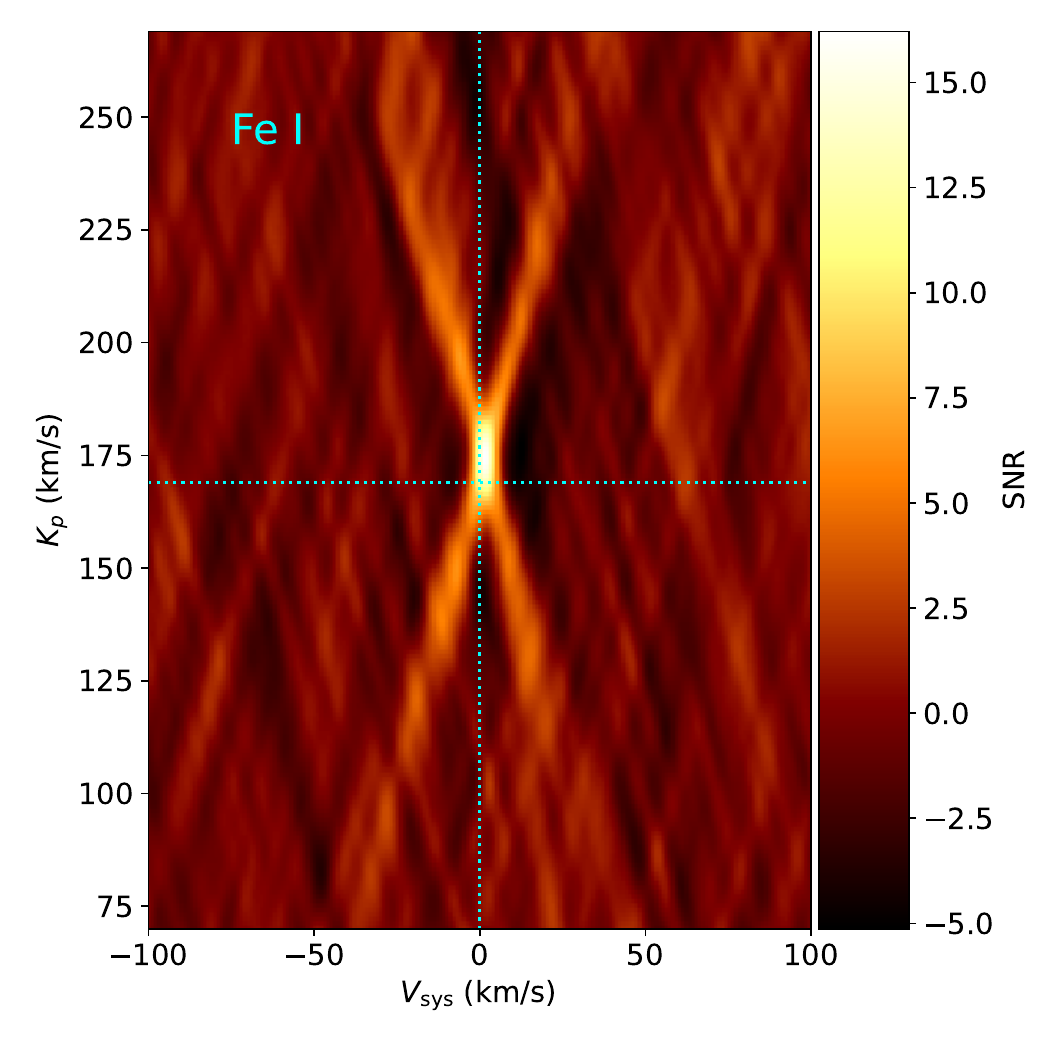}
    \includegraphics[width=\columnwidth]{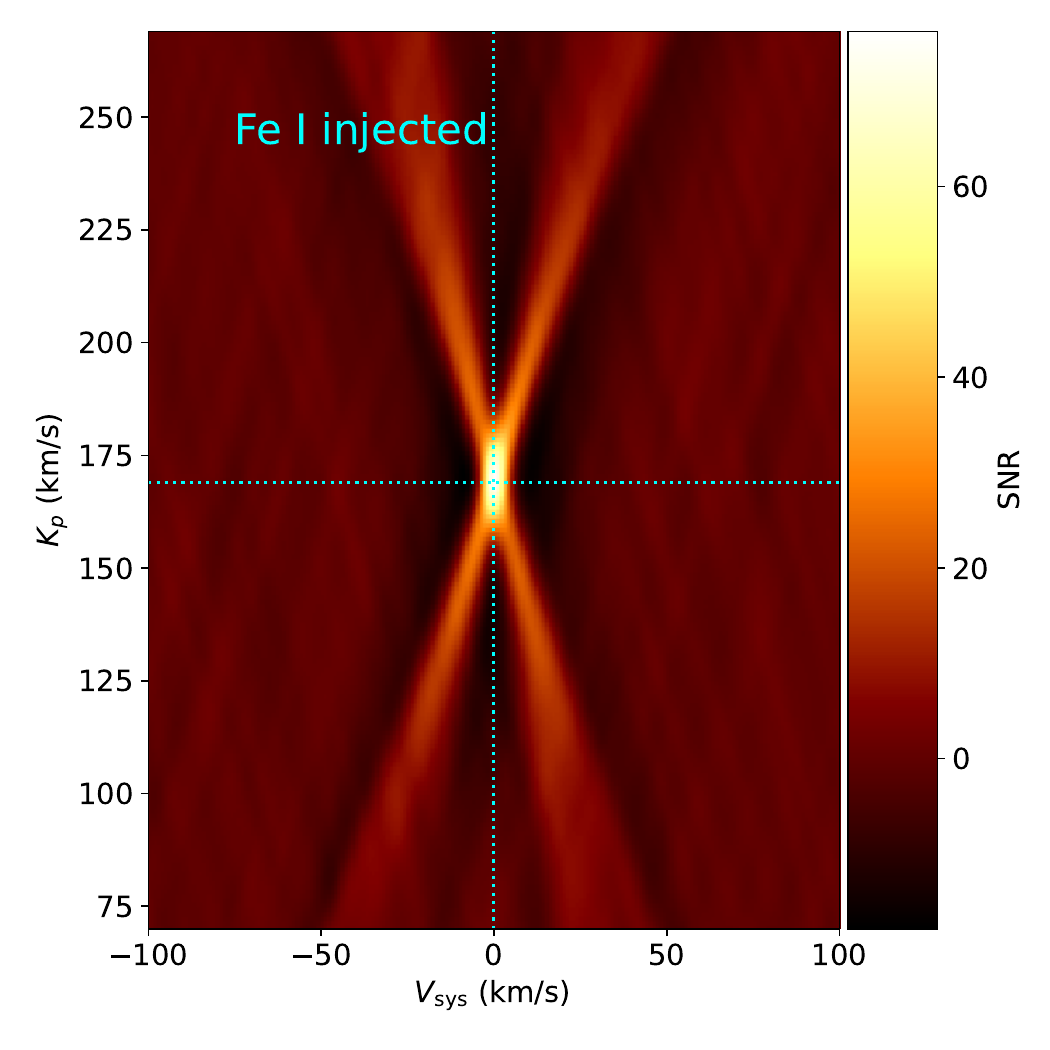}
    \\
    \includegraphics[width=\columnwidth]{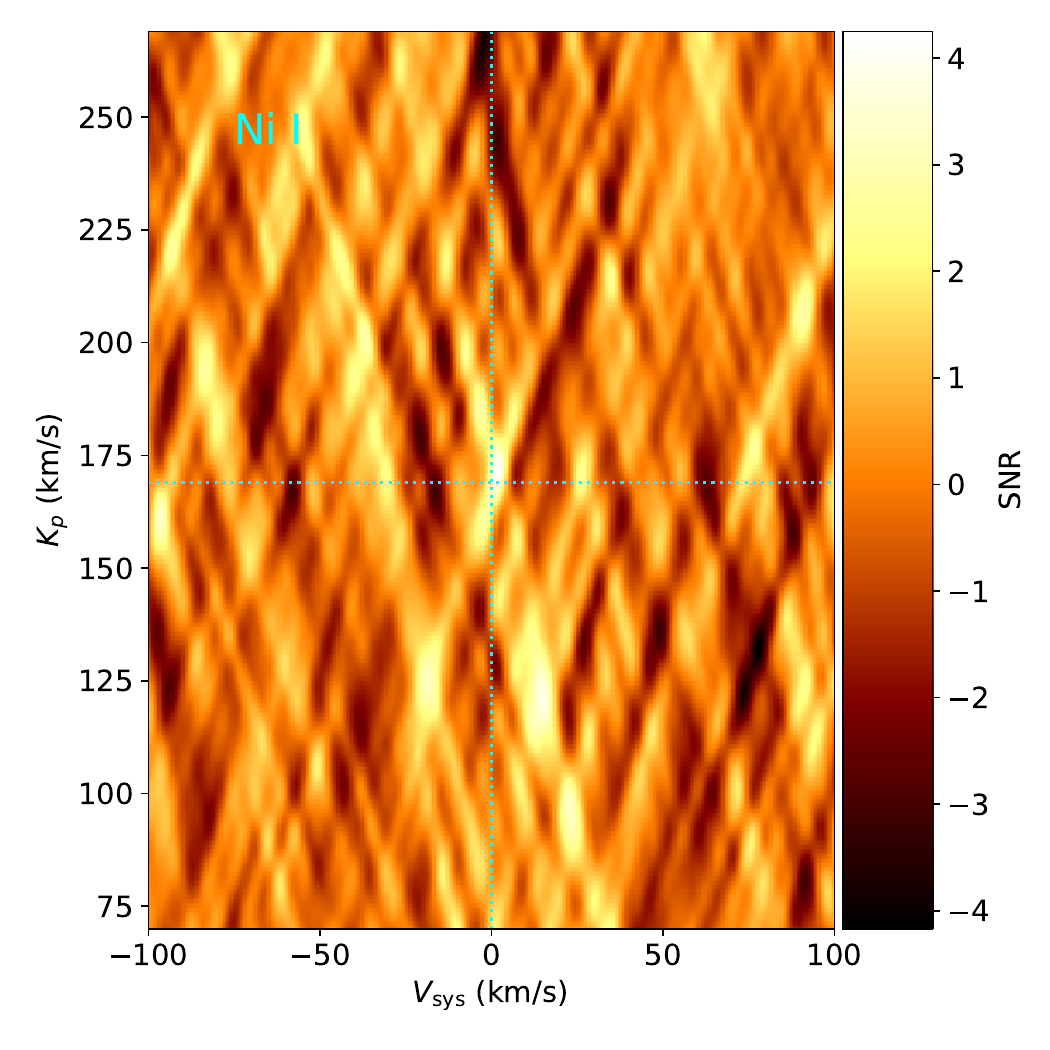}
    \includegraphics[width=\columnwidth]{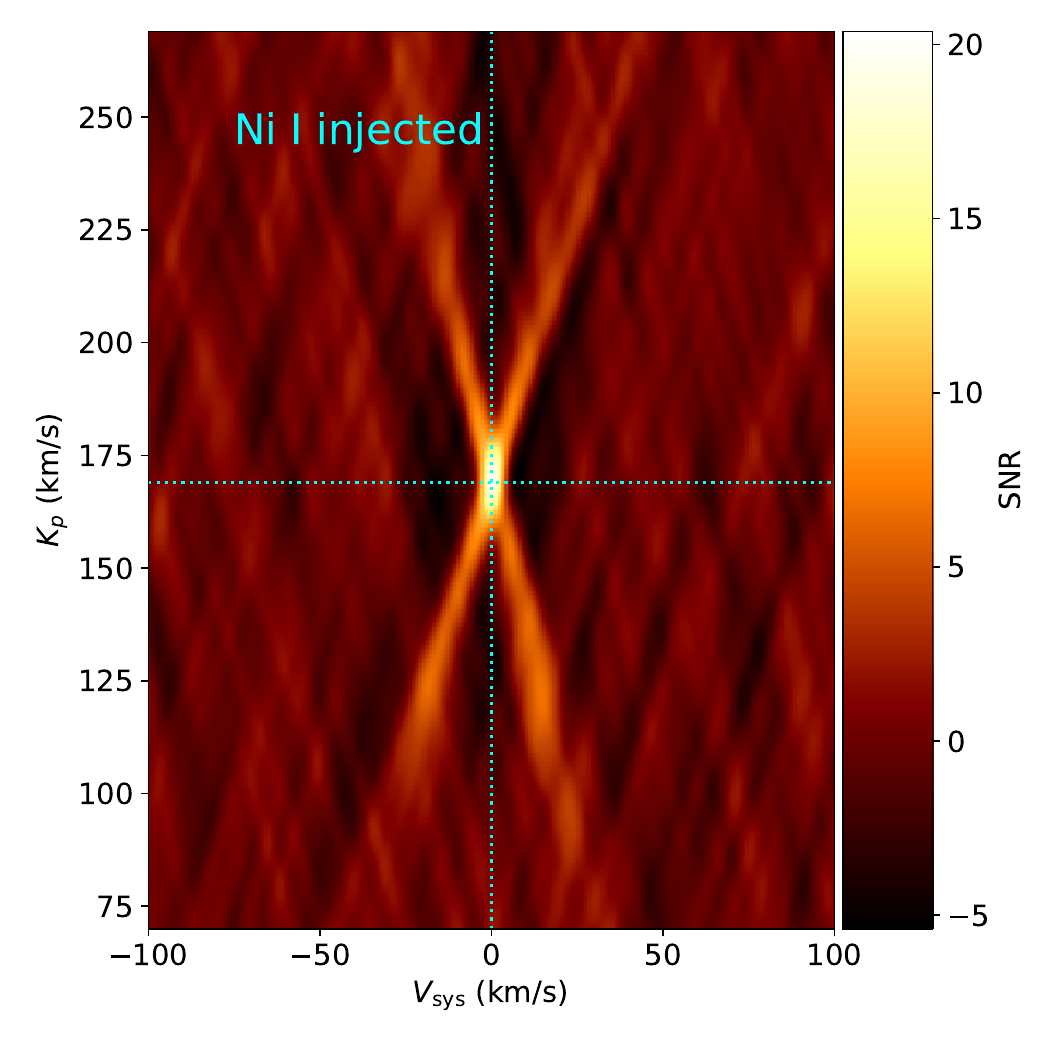}
    \caption{Top: shifted and combined CCF and injection-recovery test for Fe~\textsc{i}. The maximum SNR value measures at 17$\sigma$. Bottom: same, but for Ni~\textsc{i}. This 4.3$\sigma$ detection falls within our tentative detection range. The vertical and horizontal dashed lines in all CCF maps represent the $K_p$ and $v_{sys}$ parameters for which we should expect to find a signal.}
    \label{fig:FeICCF}

\end{figure*}

After completing a cross correlation analysis on the atomic and molecular species outlined in Figure \ref{PeriodicTable}, we were able to reproduce the 17$\sigma$ detection of Fe~\textsc{i} from \cite{Johnson2022} as seen in Figure \ref{fig:FeICCF}.
This confirmed the previous Fe~\textsc{i} emission detections from \cite{Borsa2022} and \cite{Kasper2022} and is also consistent with the detections of Fe~\textsc{i} in transmission by \cite{Casasayas2019}, \cite{Nugroho2020}, \cite{BelloArufe2022}, and \cite{Gandhi2023}.

Unlike other previous literature \citep{Cont2021a, Borsa2022}, we were not able to reproduce the detections of Cr~\textsc{i}, Fe~\textsc{ii}, or Si~\textsc{i} where we found SNRs of 4$\sigma$ or less, as seen in Figure \ref{fig:non-detections}.
Though we searched for all species outlined in Figure \ref{PeriodicTable}, we only made a clear detection of Fe~\textsc{i}. We discuss the possible reasons for these non-detections more thoroughly in \S\ref{sec:nondetectionanalysis}.

\begin{figure*}
    \includegraphics[width=0.4\textwidth]{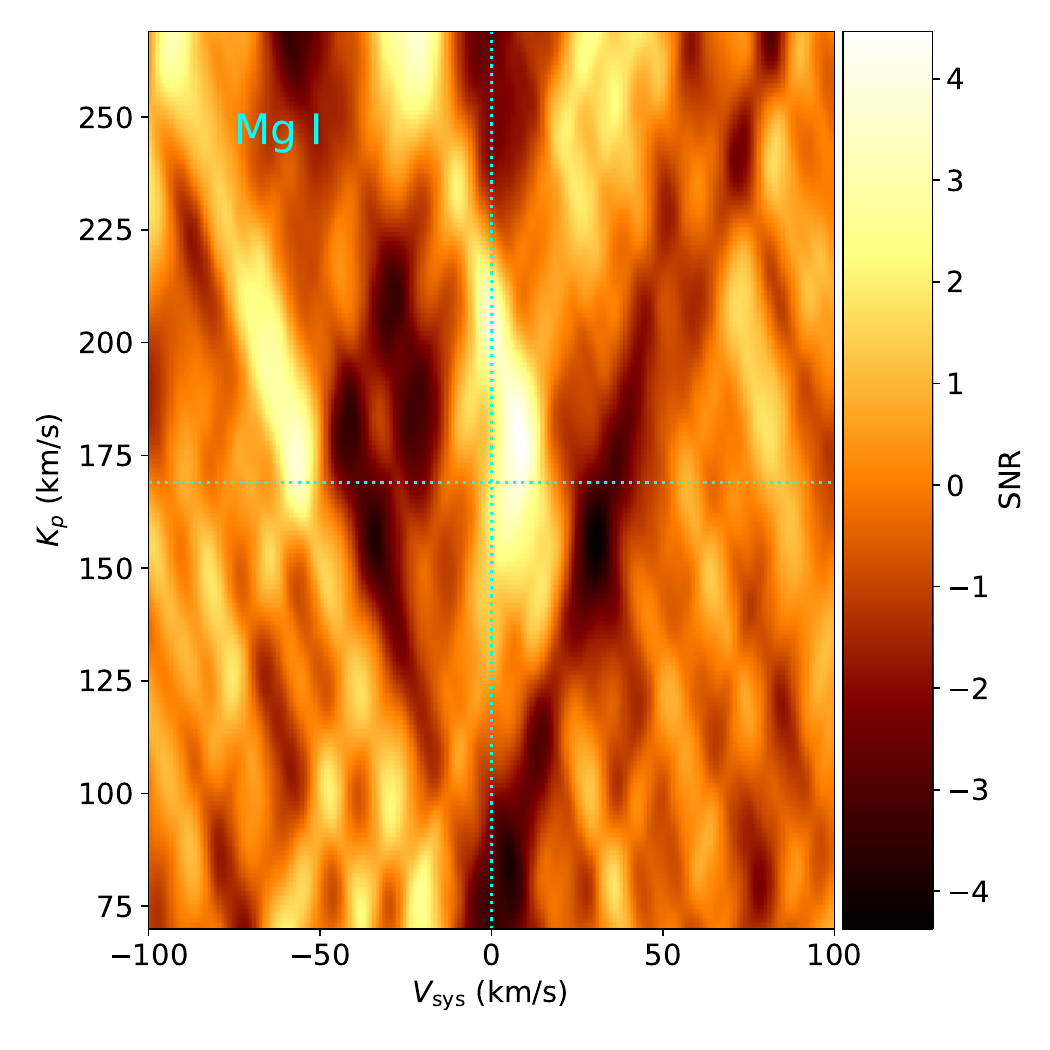} 
    \includegraphics[width=0.4\textwidth]{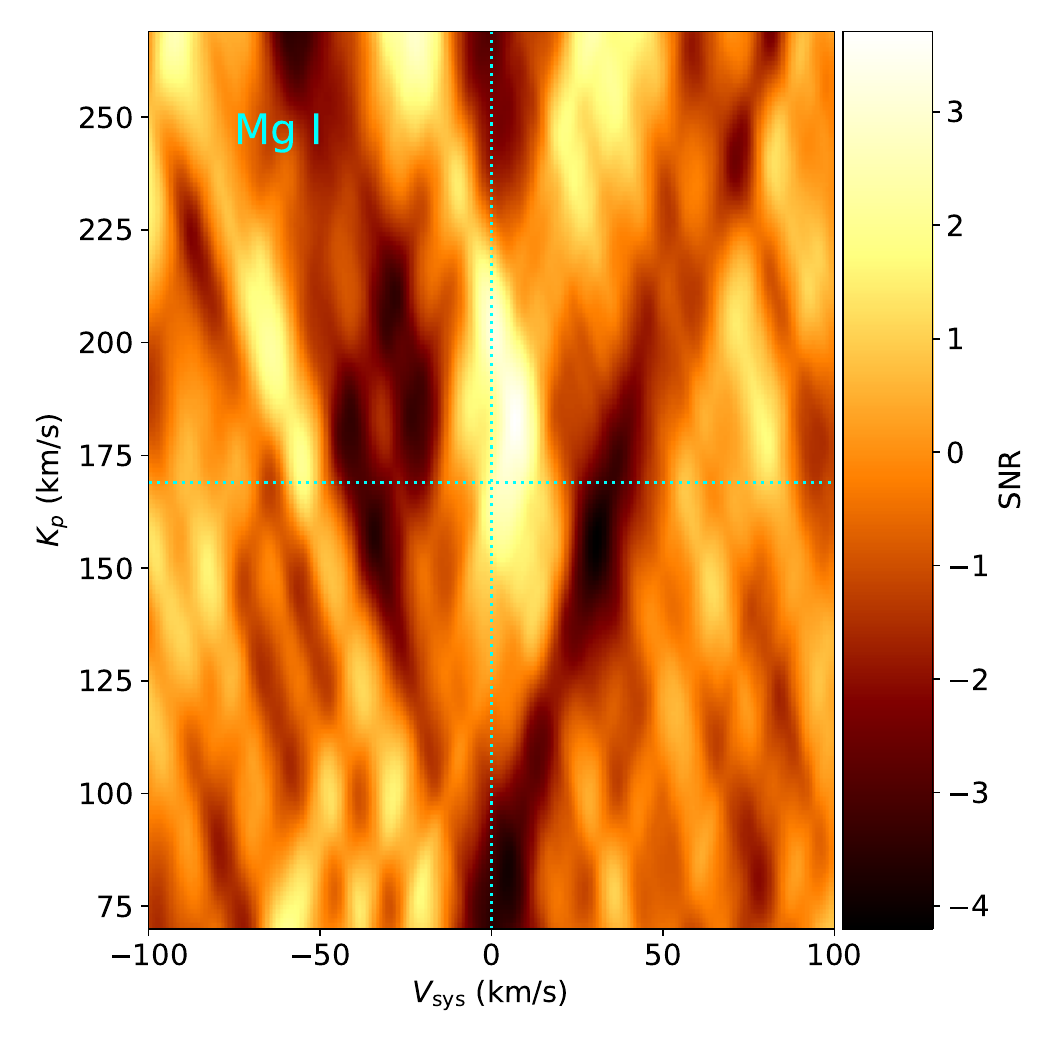} 
    \\ 
    \includegraphics[width=0.4\textwidth]{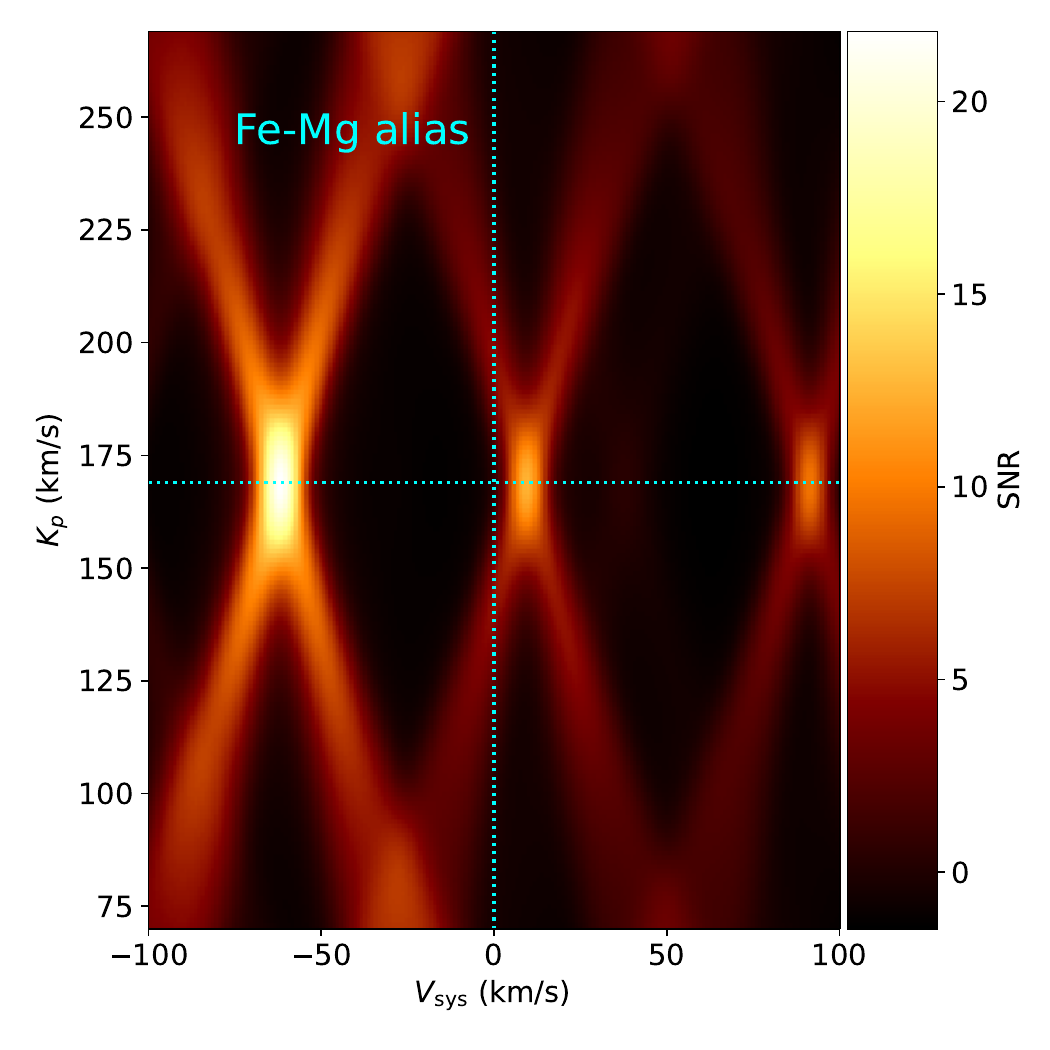}
	\includegraphics[width=0.4\textwidth]{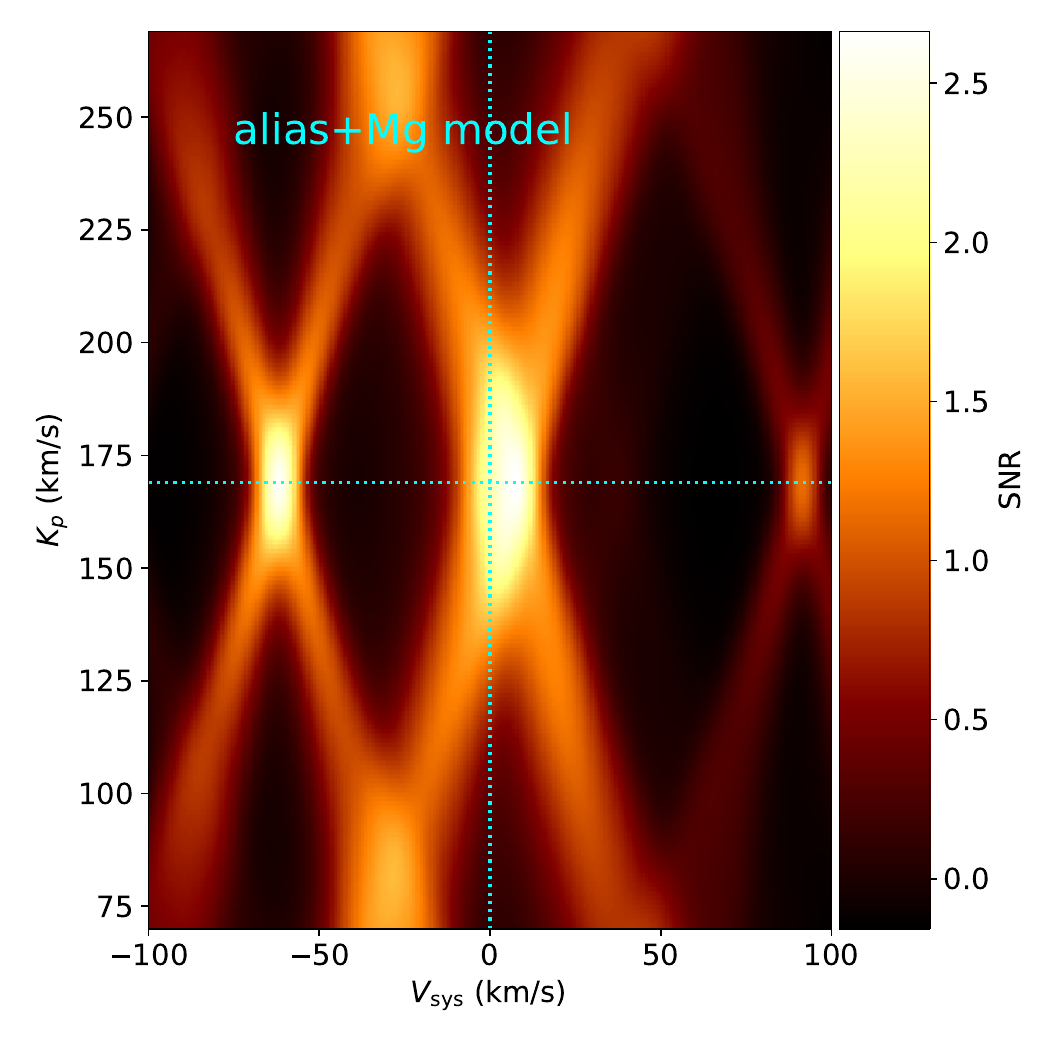}
 \\
	
    \caption{Mg~\textsc{i} CCF maps. Top: CCFs from the data, showing a peak near $v_\mathrm{sys}=0$ km s$^{-1}$ and a second peak near $v_\mathrm{sys}=-50$ km s$^{-1}$. Bottom row: alias patterns caused by cross-correlating an Mg~\textsc{i} template spectrum with a model spectrum containing (left) only Fe~\textsc{i}, and (right) both Fe~\textsc{i} and Mg~\textsc{i}. In both cases, the model spectra have been shifted according to the expected planetary RV at each observation epoch, cross-correlated with the template, and then shifted back into the planetary rest frame. The overall alias pattern is similar to that seen in the data.}
    \label{fig:MgI}
\end{figure*}

Although we only make one clear detection of a species, we find that Ni~\textsc{i} (Fig.~\ref{fig:FeICCF}) and Mg~\textsc{i} (Fig.~\ref{fig:MgI}) fall within our tentative detection range, with SNRs of 4.3$\sigma$ and $4.5\sigma$, respectively. Both signals deserve further investigation. 

\begin{figure*}
    \includegraphics[width=0.4\textwidth]{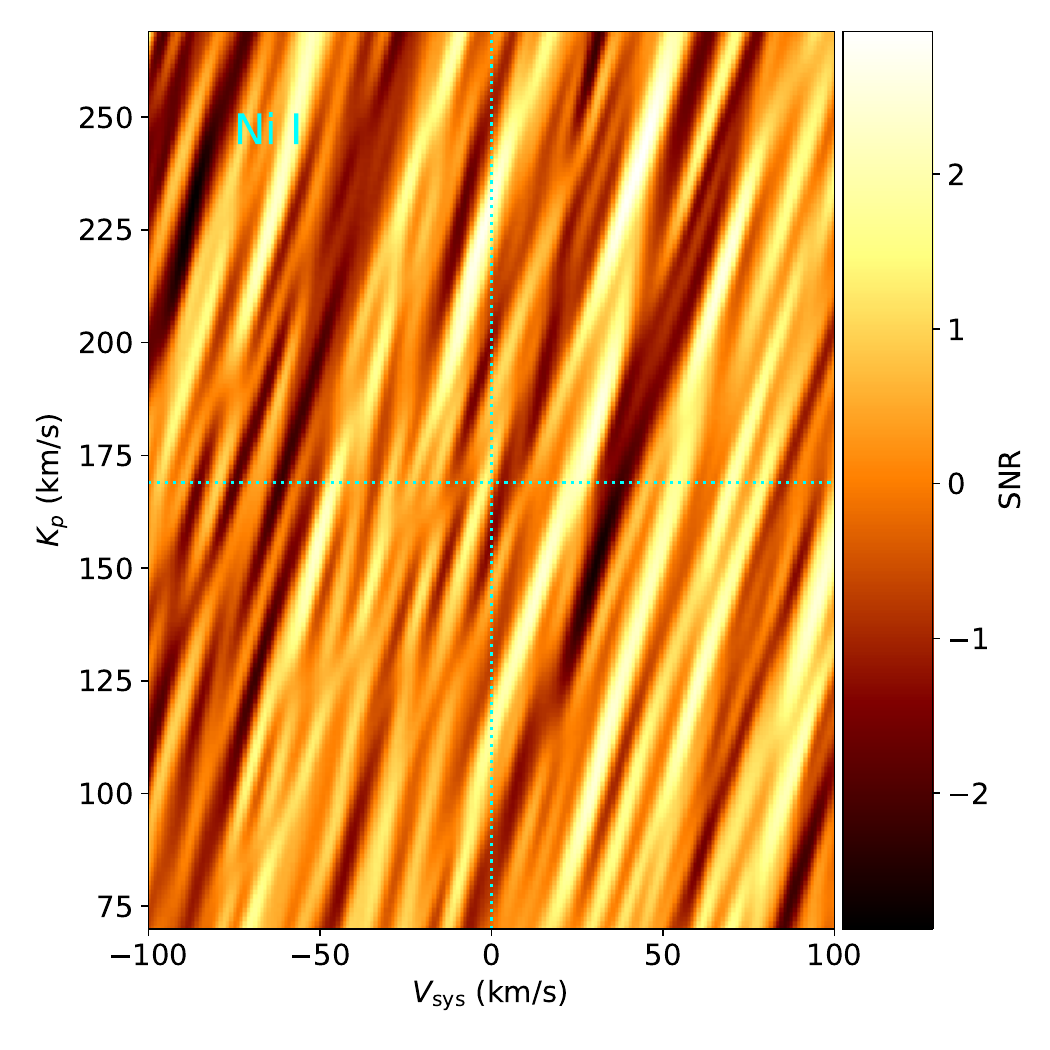}\includegraphics[width=0.4\textwidth]{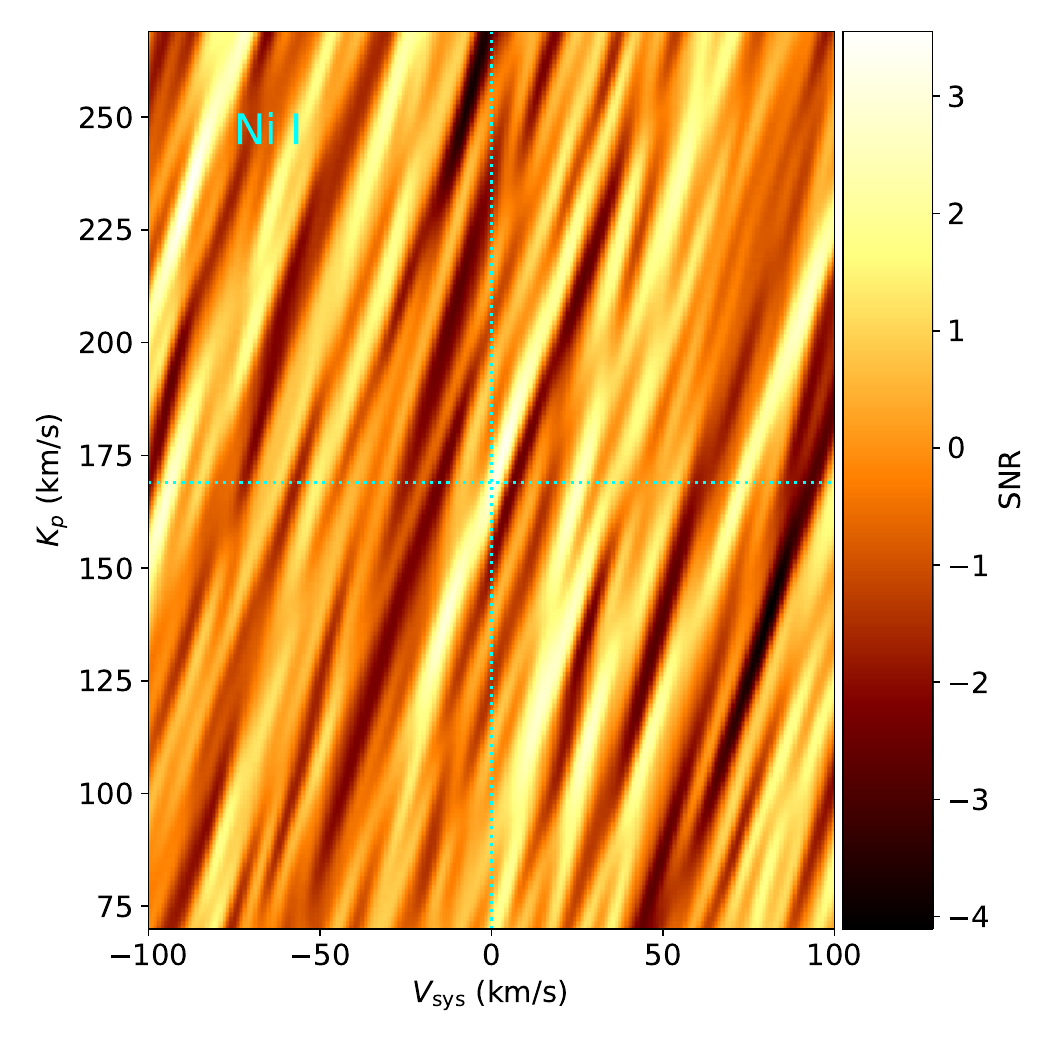} \\ 
	\includegraphics[width=0.4\textwidth]{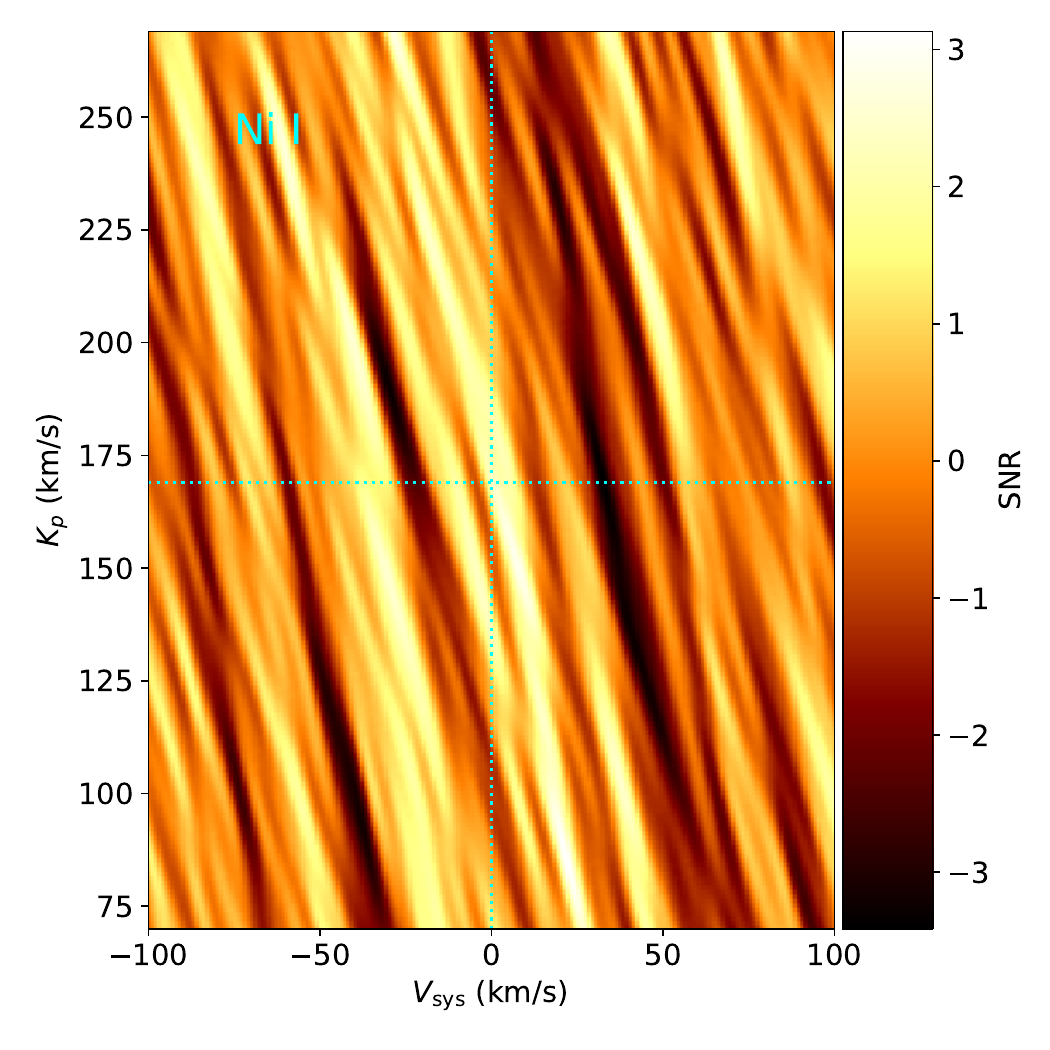}\includegraphics[width=0.4\textwidth]{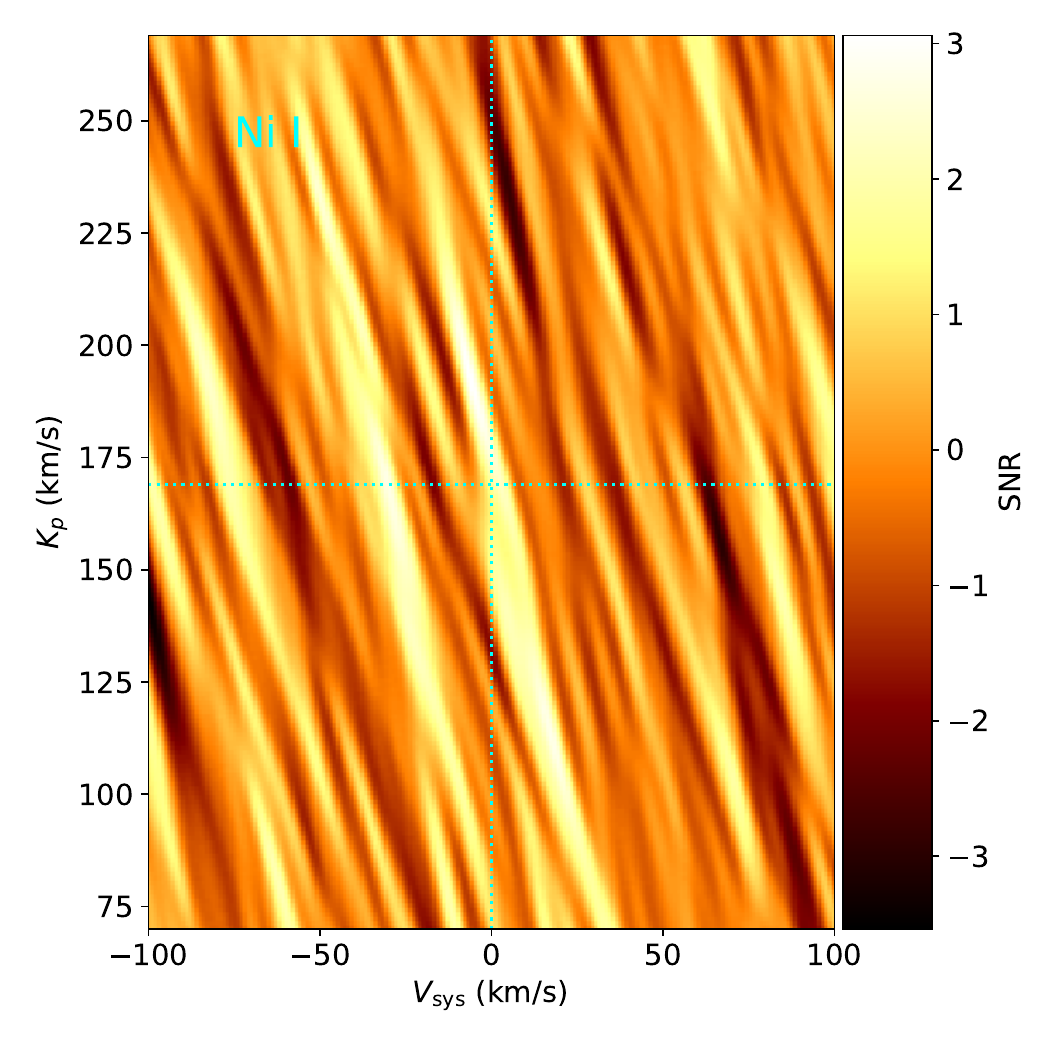} \\
	
    \caption{Ni~\textsc{i} CCF maps before their combination. Maps in the left column were created using data from PEPSI's blue arm, and maps in the right column were created using data from PEPSI's red arm. The top row displays maps with data taken on May 1st, 2022 and the bottom row displays maps with data taken on May 18th, 2022.}
    \label{fig:NiI}
\end{figure*}

For Ni~\textsc{i} we look at the peaks on the CCF maps for each PEPSI arm and each night of observation as seen in Figure \ref{fig:NiI}.
We find that there are stronger peaks in both arms on our second night of observation and that we only see a peak in the blue arm on our second night of observation.
Though for a clear detection, we would expect to see strong peaks in each arm and on both observation periods, most Ni~\textsc{i} lines are present in the red part of the optical, and therefore we might expect a stronger signal in the red. We cannot either confirm or refute the Ni~\textsc{i} signal with the current data, but additional data could allow for a more definitive result.

Mg~\textsc{i} has not been previously detected in emission for KELT-20b, although it has been detected in transmission \citep{Gandhi2023}. 
We confine our analysis to the PEPSI blue-arm data, as Mg~\textsc{i} has several strong lines in CD III and only a handful of weak lines in CD V (Fig.~\ref{ModelSpectra}). Although these data do display a prominent peak near the expected location, another peak is evident with $v\sim-55$ km s$^{-1}$. This pattern is likely due to aliasing with Fe~\textsc{i}.

As described in detail by \cite{Borsato2023}, species with few lines (for instance, Mg~\textsc{i}) can alias with species with many lines (for instance, Fe~\textsc{i}) to produce spurious cross-correlation peaks. In Fig.~\ref{fig:MgI}, we show the alias structures produced by cross-correlating our Mg~\textsc{i} template with the Fe~\textsc{i} template, and the sum of the Fe~\textsc{i} and Mg~\textsc{i} template. The pattern of the alias structure in the $v_\mathrm{sys}-K_P$ plot is a good qualitative match for that seen in the data (top left panel of Fig.~\ref{fig:MgI}). 
Inspired by \cite{Borsato2023}, we fit the $-55$ km s$^{-1}$ peak with the Fe~\textsc{i}-Mg~\textsc{i} alias model, with the sole free parameter scaling the amplitude of the alias signal. We perform this fit to the one-dimensional CCF at the nominal $K_P$ of KELT-20b ($K_P=169$ km s$^{-1}$), and downsample the CCF to a grid spacing of 3 km s$^{-1}$ (similar to the velocity resolution of PEPSI) in order to avoid correlations between adjacent grid points. We then subtract off this scaled alias model from the two dimensional CCFs, as shown in Fig.~\ref{fig:MgI}. This suggests that the greater part of the peak near 0 km s$^{-1}$ is due to the Fe~\textsc{i} alias, and the remaining power is insufficient to claim a detection. Nonetheless, much like with Ni~\textsc{i}, additional data and/or a better treatment of the Fe~\textsc{i} alias could potentially increase the signal and allow for a more confident detection of Mg~\textsc{i} emission. 

Additionally, we find a small redshift of $\sim4$ km s$^{-1}$ of the Fe~\textsc{i} and tentative Ni~\textsc{i} and Mg~\textsc{i} signals. This is expected due to the global-scale day-to-nightside winds, which will introduce a net redshift on the emission lines. A blueshift of $\sim2-6$ km s$^{-1}$ has previously been found from transmission spectroscopy measurements \citep{Casasayas2019,Nugroho2020,Stangret2020,Pai2022}. A full analysis of the atmospheric dynamics is beyond the scope of the present work but will be presented in a future paper.

\subsection{Assessment of non-detections}
\label{sec:nondetectionanalysis}

To investigate why these results may differ from previous literature, we remove one variable that could cause our lack of detections - the P-T profile. Changes in the P-T profile can change the relative line strengths, which could affect the strength of the recovered signal; for a weak signal, this could potentially cause a detection to drop beneath the detection threshold. For example, \cite{Johnson2022} demonstrated that adopting various literature P-T profiles for KELT-20b resulted in Fe~\textsc{i} detection significances ranging from 13 to 17$\sigma$. 
To test this, we adopt P-T profiles from \cite{Kasper2022, Borsa2022} and \cite{Yan2022b}, the last used in \cite{Cont2021a}. We show these profiles in Fig. ~\ref{fig:ptprofile}.
After repeating our methodology with the above profiles, none of these tests resulted in detections. We also note that the \cite{Johnson2022} tests referenced above indicated that the Guillot profile that we adopted earlier in this work resulted in the strongest Fe~\textsc{i} detection.

\begin{figure*}
    \includegraphics[width=0.4\textwidth]{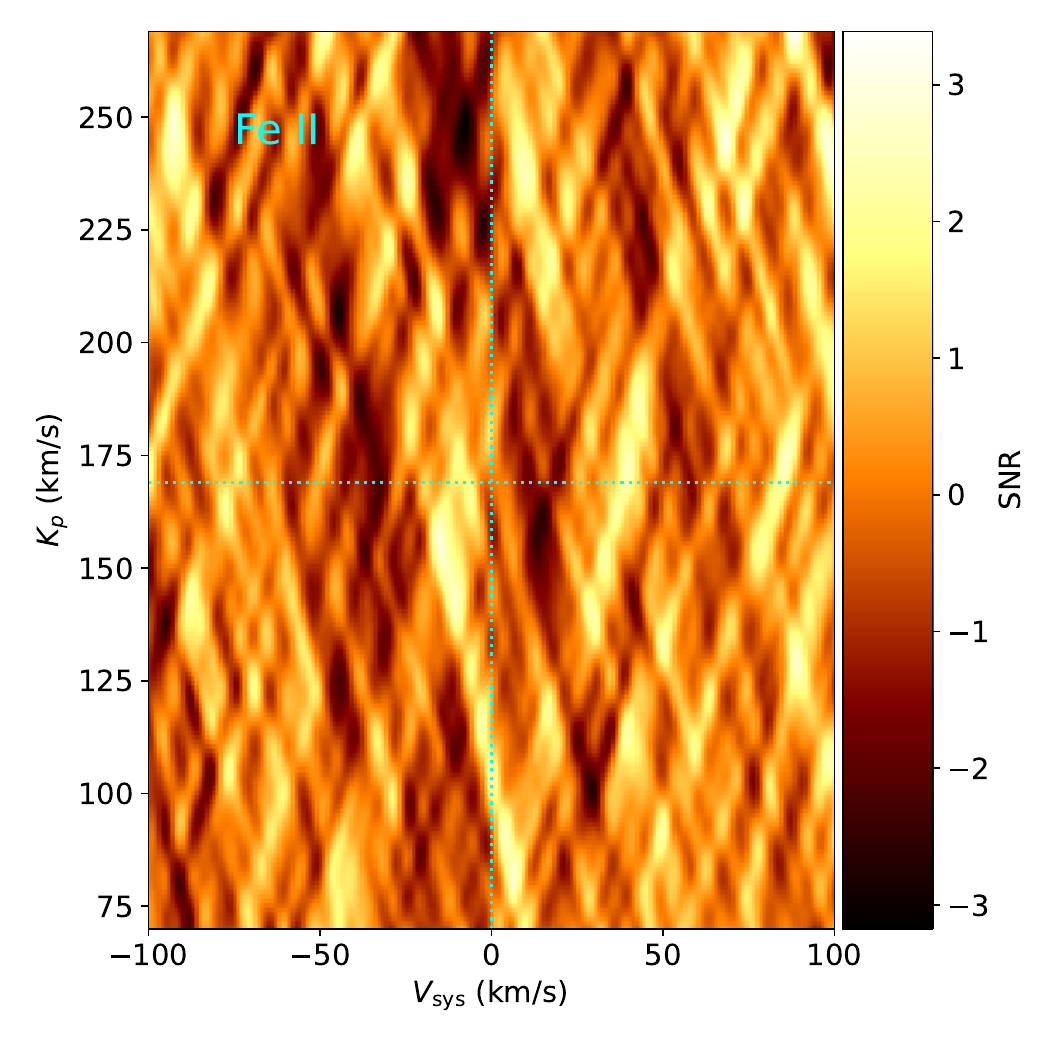}\includegraphics[width=0.4\textwidth]{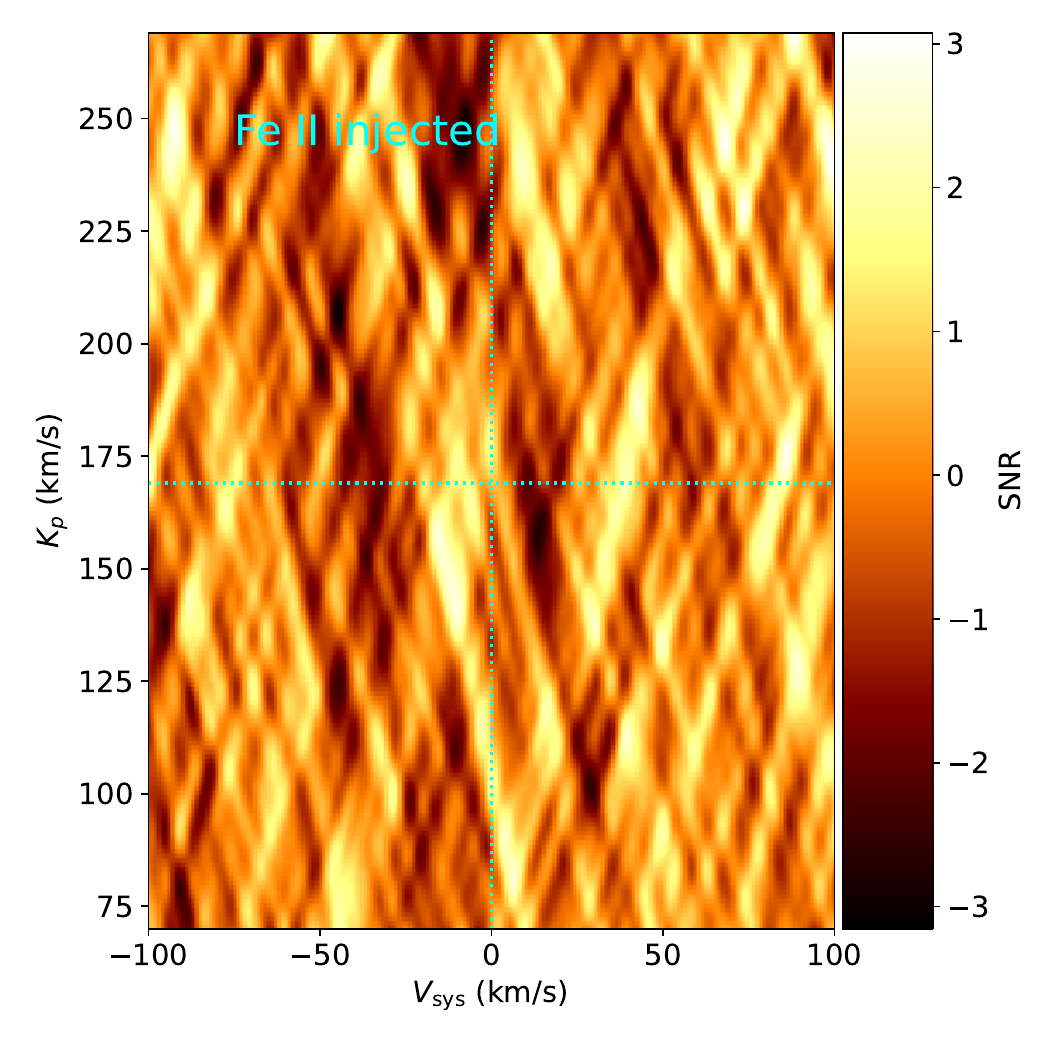} \\
    \includegraphics[width=0.4\textwidth]{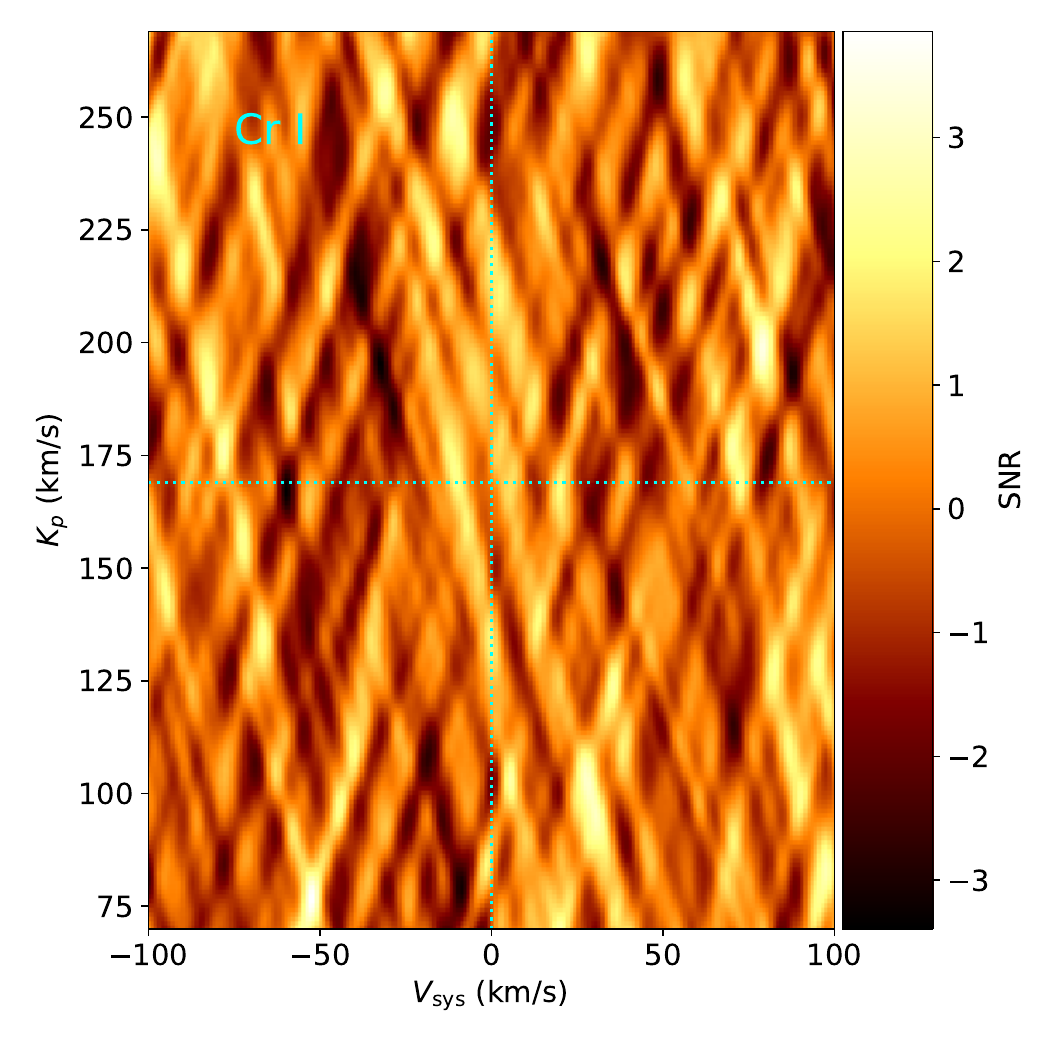}\includegraphics[width=0.4\textwidth]{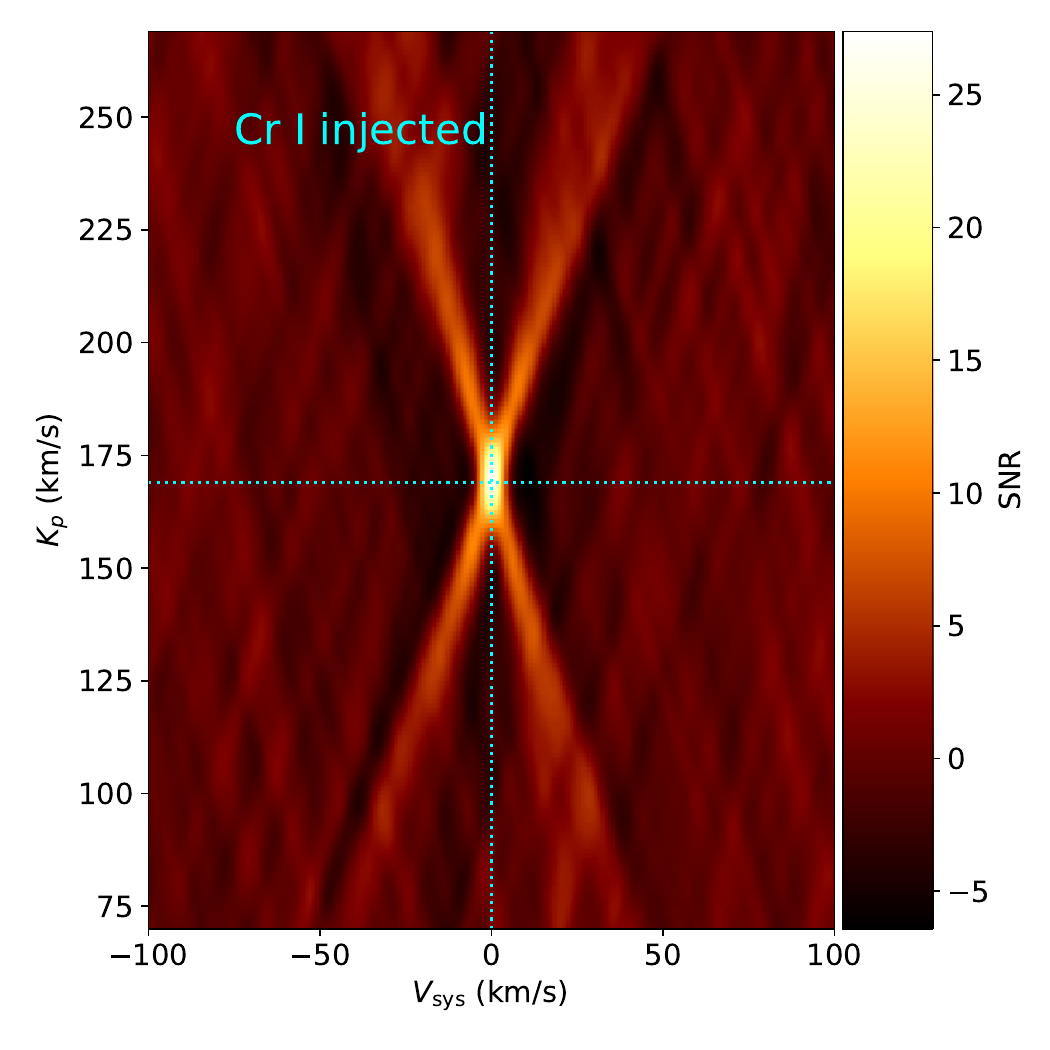} \\
    \includegraphics[width=0.4\textwidth]{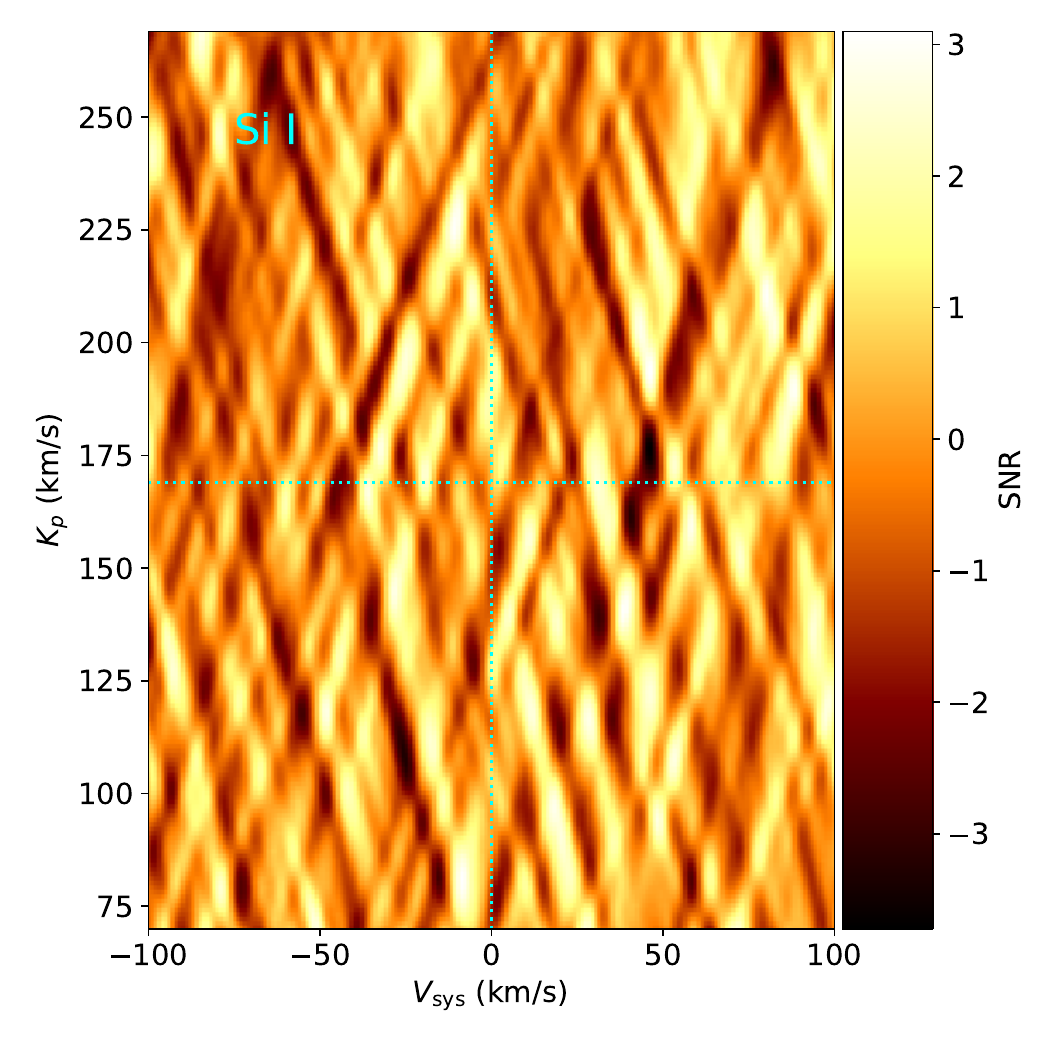}\includegraphics[width=0.4\textwidth]{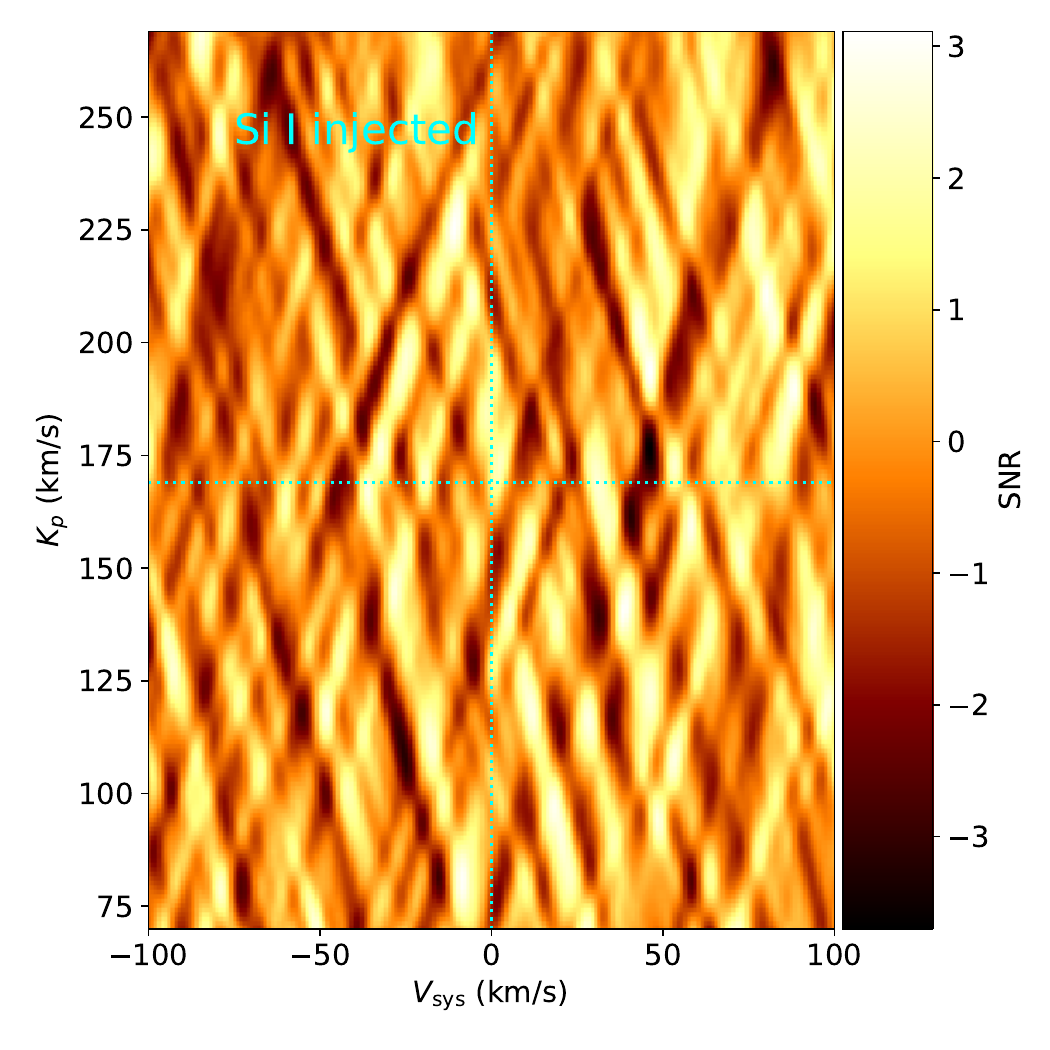} \\
    \caption{Shifted and combined CCF maps and injection-recovery tests for Fe~\textsc{ii}, Cr~\textsc{i}, and Si~\textsc{i}. Unlike Fe~\textsc{i}, these species have no significant detection. Additional notable non-detections are shown in Figure \ref{fig:additionalnon-detections}.}
    \label{fig:non-detections}
\end{figure*}

To further explore why our results may differ, we performed injection-recovery tests as outlined in Section \ref{subsection:CCF}. 
By implementing these tests, we can determine if our inability to reproduce previous results is due to our data or our atmospheric models. 
After running injection-recovery tests on the species outlined in Table \ref{tab:testedspecies}, we find that we can detect each of these species (>4$\sigma$) except for Fe~\textsc{ii} and Si~\textsc{i} (Si~\textsc{i} and Cr~\textsc{i} shown in Figure \ref{fig:non-detections}).
This result indicates that species detected using this test should have been detected with a high significance (recovered SNR reported in Table \ref{tab:testedspecies}), implying that their concentration in the atmosphere is different that we would expect given our P-T profile, and that our non-detections are not the result of insufficient data quality.
We do note that these tests are overly optimistic, as the injected model and CCF template spectra are identical; for instance, for Fe~\textsc{i}, we obtained a 17.0$\sigma$ detection while injection-recovery tests predict an SNR of 76.3$\sigma$ and for Ni~\textsc{i} predicts a 20.4$\sigma$ signal compared to an actual $4.3\sigma$ tentative detection. Nonetheless, this suggests that the SNR is likely overestimated by a factor of a few in the injection-recovery tests, not by orders of magnitude, which suggests that the predicted 27.4$\sigma$ Cr~\textsc{i} signal should have resulted in a detection.

This result also indicates that our model lines are not strong enough to predict a detection of Fe~\textsc{ii} or Si~\textsc{i}.
To re-test these species with stronger lines, we can assume that all of the element is in these species and calculate new VMR values using Equation \ref{eqn:VMR}.
Equation \ref{eqn:VMR} returns a higher VMR than given by \texttt{FastChem}, so we consider this to be the maximum VMR for these species.
When we run injection-recovery tests using these models, we still are not able to recover a detection for these species. 
We thus do not expect to be able to reproduce the previous detections of Si~\textsc{i} and Fe~\textsc{ii}, consistent with our non-detections.

We can attempt to quantify this potential difference in detectability with previous literature by repeating our methodology from Section \ref{sec:quant}, but instead finding the area under the model spectra over the wavelength range of other instruments used to produce claimed detections.
\cite{Borsa2022} uses HARPS-N, which has a wavelength range that covers a broader range of blue in the optical than PEPSI, spanning 3900-6900\AA, while \cite{Kasper2022} used MAROON-X, which has more red optical coverage (5000-9200 \AA) than PEPSI.
For both Cr~\textsc{i} with HARPS-N and Ni~\textsc{i} with MAROON-X, we compute detectability values which are a factor of $\sim2$ higher than with PEPSI.

On the other hand, the estimated total SNR for each of our datasets (SNR$_{\mathrm{mean}}\times\sqrt{N}$, where SNR$_{\mathrm{mean}}$ is the mean SNR of a dataset and $N$ is the number of spectra) is approximately twice that of the published datasets from \cite{Borsa2022} and \cite{Kasper2022}.
While the trade-off between detectability (equivalently, bandpass) and SNR is likely to be complex and non-linear, these suggest that it is not obvious that the published HARPS-N dataset is significantly more optimal than PEPSI for the detection of Cr~\textsc{i}, nor MAROON-X for the detection of Ni~\textsc{i}.

Si~\textsc{i} was previously detected with CARMENES, a near-infrared spectrograph \citep{Cont2021a}; this species has many more lines in the near-IR than the optical, so it is not unsurprising that we are unable to detect Si~\textsc{i}. 

The Fe~\textsc{ii} non-detection is more surprising, as it does have some significant lines in the optical (Fig.~\ref{ModelSpectra}); indeed, it has been detected in transmission for this same planet using this same instrument setting \citep{Pai2022}. We can only conclude that this is due to the greater sensitivity of transmission spectroscopy than emission spectroscopy \citep[e.g.,][]{Johnson2022}. There are additional lines between 4000 and 5000 \AA\ captured by HARPS-N but not PEPSI (Fig.~\ref{ModelSpectra}) and used by \cite{Borsa2022}. 
These additional lines give HARPS-N a detectability value $\sim2$ times that PEPSI.
It is unclear whether the additional lines are enough to compensate for the lower SNR of the HARPS-N dataset as compared to PEPSI; a complete assessment of this issue is potentially complicated and beyond the scope of this paper.

We also note that the removal of a large number of systematics with SYSREM can distort or attenuate the planetary signal \citep[e.g.,][]{Nugroho2017} and could potentially result in a false negative. We, however, are removing no more than three systematics, significantly fewer than the 10-15 used by some previous works on other planets \citep[e.g.,][]{Nugroho2017,Gibson2020}, which should not significantly attenuate the signal.

\cite{Borsa2022} found signals of Cr~\textsc{i} and Fe~\textsc{ii}, but only detected them on one side of KELT-20b's secondary eclipse. 
They invoked changing line strengths as a function of phase to explain their observations, as has been observed and predicted for other planets \citep{vanSlujis2022, Herman2022, Beltz}. 
We tested for this by considering each night of our observations alone, and still could not detect either Fe~\textsc{ii} or Cr~\textsc{i} emission.
Furthermore, the \cite{Borsa2022} detections fall below our conservative detection limit ($3.9\sigma$ for Fe~\textsc{ii} and 3.6$\sigma$ for Cr~\textsc{i}).
\cite{Kasper2022} were also unable to reproduce the detections of Cr~\textsc{i}, Fe~\textsc{ii}, or Si~\textsc{i}, although they did not carry out a quantitative assessment of whether they should have been able to detect these species. 

Overall, we conclude that our non-detections of Fe~\textsc{ii} and Si~\textsc{i} are consistent with expectations from these species' spectra and our data, while our tentative detection of Ni~\textsc{i} is consistent with the stronger detection from \cite{Kasper2022} and the smaller number of Ni~\textsc{i} lines in the PEPSI bandpass. We expect, however, that we should have detected the Cr~\textsc{i} signal found by \cite{Borsa2022}, as well as from the expected Cr~\textsc{i} concentration in the planetary atmosphere.

\section{Conclusions}
We have presented an analysis of high-resolution PEPSI emission spectra of the UHJ KELT-20b.
After examining a large variety of atomic and molecular species we detect Fe~\textsc{i} with a significance of 17$\sigma$, a potential 4$\sigma$ signal from Ni~\textsc{i}, and do not clearly detect any other species. 
While we also detect a $4\sigma$ Mg~\textsc{i} signal from the blue arm of PEPSI, this signal can be partially explained by an alias with the Fe~\textsc{i} lines already known to be present in the spectrum. After removal of the alias signal, the residual Mg~\textsc{i} signal does not meet our detection threshold.

Our detection of Fe~\textsc{i} allows us to come to the same conclusion as previous literature \citep{Johnson2022, Yan2022a, Borsa2022}, that the spectrum is dominated by Fe~\textsc{i}, demonstrating its presence in the atmosphere.
Fe~\textsc{i} appears to by the most significant source of line capacity in the optical, giving it a high likelihood to be at least partially responsible for KELT-20b's temperature inversion, especially given the non-detections of molecular inversion agents from this work and \cite{Johnson2022}.

Though it is possible that Fe~\textsc{i} alone may be responsible for this temperature inversion \citep{Lothringer2018}, we cannot rule out continuum opacity sources such as H$^-$, which can occur in hot planetary atmospheres \citep{Arcangeli2018}.
Because high-resolution spectroscopy lacks the ability to see wider spectral features due to the data reduction process used to perform a CCF methodology, H$^-$ may be practically invisible. 
This data reduction process normalizes the continuum, removing any broadband features of the spectrum.
Though we are not able to see wider spectral features, they still may be contributing to the temperature inversion on KELT-20b. 
Future use of low-resolution spectroscopy may be able to constrain the presence of H$^-$ or other continuum opacity sources.

We find a tentative detection of Ni~\textsc{i} with a significance of 4.3$\sigma$, backing up previous findings from \cite{Kasper2022}. 
While we are unable to completely confirm this detection as it falls within our tentative detection range of 4-5$\sigma$, this result gives us reason to further investigate Ni~\textsc{i} with future observations. An initially promising $4.5\sigma$ Mg~\textsc{i} detection is contaminated by an Fe~\textsc{i} alias; when corrected a CCF peak is still present near the expected location for a true signal, but falls below our $4\sigma$ detection threshold.

Other than Fe~\textsc{i}, Ni~\textsc{i}, and Mg~\textsc{i}, we searched for 64 other atomic and molecular constituents in the atmosphere of KELT-20b and did not make any other detections.
We did not detect the additional species previously detected in transmission.
Transmission spectroscopy probes a higher altitude of the atmosphere and near the day/night boundary than emission spectroscopy, and our previous work with these datasets suggested that we are more sensitive to trace species with transmission spectroscopy \citep{Johnson2022}. Given these differences, it is not necessarily surprising that we do not detect the other species detected in transmission, which could be due to a combination of sensitivity and atmospheric inhomogeneity. 
Notably we did not detect Si~\textsc{i}, Cr~\textsc{i}, and Fe~\textsc{ii} despite their claimed detections with emission spectroscopy in other literature \citep{Cont2021a, Borsa2022}.
While our injection-recovery tests indicate that with our data we should not have been able to detect Si~\textsc{i} or Fe~\textsc{ii} due to the small number of lines in the PEPSI bandpass, our non-detection of Cr~\textsc{i} is unexpected. The smaller number of Cr~\textsc{i} lines within the PEPSI bandpass as compared to HARPS-N should have been at least approximately compensated by our higher SNR data, while our injection-recovery tests suggest that we should have been able to detect Cr~\textsc{i}. 
This perhaps suggests that the concentration of Cr~\textsc{i} in the atmosphere of KELT-20b is different that we would expect given we assumed a constant VMR as a function of altitude, which is a simplistic assumption.

A proposed potential reason for these differences is rain-out in the atmosphere \citep{Visscher2010, Parmentier2013}. We assume that KELT-20b is tidally locked in its orbit, which causes the atmosphere to heat up on its day side, and cool on its night side, causing some species to condense and fall into lower levels in the atmosphere where they are harder to detect. Due to this effect, species that we otherwise expect may not be detectable.

Another potential reason for a difference in results could be the presence of general atmospheric variability in KELT-20b. 
UHJs have large temperature differences between their day and night sides, resulting in global day-to-nightside winds and superrotating equatorial jets.
Though there is no solid evidence of variability in KELT-20b \citep{Pai2022}, there is some evidence for atmospheric varibility in other similar hot planets.
Some of which include variable winds in KELT-9b \citep{Pai2022}, potential cloud changes on hot brown dwarf KELT-1 \citep{Parviainen2023}, and changes in spectral strength in UHJ WASP-121b \citep{Qinglin2023}. 
Atmospheric variability could in principle result in a Cr~\textsc{i} detection at only some epochs, but between our datasets and those of \cite{Borsa2022} and \cite{Kasper2022}, this only resulted in a signal for one out of six epochs.

This variability, as well as general chemical disequilibrium processes are difficult to account for in atmospheric modeling.
Due to this difficulty, the \cite{Guillot} P-T profile, as well as profiles from \cite{Kasper2022, Borsa2022} and \cite{Yan2022b} do not account for these factors.
In order to make better constraints on chemical constituents in the atmosphere of KELT-20b, it will be necessary to utilize modeling that more accurately portrays these atmospheric processes.

\subsection{A need for 
repeatable analysis}
Because of the differences between recent results regarding the atmospheric chemistry of KELT-20b, it's important to perform 
a homogeneous analysis using high resolution spectroscopy and a high SNR data set to attempt to resolve these discrepancies.
Discrepancy between results using high resolution spectroscopy to study exoplanet atmospheres has occurred from the earliest days of the field, starting with a claimed detection of reflected starlight from $\tau$ Boo b in \cite{Collier99}.
However, simultaneous work from \cite{Charbonneau99} failed to detect the proposed signal of a highly reflective $\tau$ Boo b using a similar methodology, instead only presenting an upper limit.
With updated data, reanalysis using high resolution optical spectra was completed in \cite{Collier2000}, which was unable to replicate the same feature found in their previous work, but further constrained previously reported upper limits for the albedo of $\tau$ Boo b.
Using improved high resolution Doppler techniques, \cite{Leigh2003} utilized new data as well as further re-analyzed the original data from \cite{Collier99}.
Though their work suggested a weak candidate signal at the most probable radial velocity amplitude, they concluded that its statistical significance was too weak to claim a detection of any strength. More recent searches for reflected starlight have also failed to find anything more than tentative signals \citep{Rodler2010,Rodler2013}. It was not until over a decade after the initial claims of reflected starlight that $\tau$ Boo b was detected via optical Doppler spectroscopy, but using CO absorption in the infrared, not reflected light in the optical \citep{Brogi2012}.
Previous detections of reflected light from other planets, such as 51 Peg b, have also been unable to be replicated with optical high-resolution spectroscopy \citep{Scandariato2021}.

This discrepancy between results is still 
not uncommon in high resolution UHJ observations, as mentioned in Section \ref{section:intro} in regards to multiple molecular species.
\cite{Nugroho2017} and \cite{Cont2021a} claimed detections of TiO in transmission and emission in WASP-33b, however these results could not be replicated by other literature including \cite{Herman2020} and \cite{Serindag2021}, who suggest that the choice of line lists and P-T profiles to be a strong contender for the inconsistencies between results.
This is further emphasized in \cite{Merritt2020}, who reported a non-detection of VO in WASP-121b, attributing the differences from \cite{Hoeijmakers2020} to be related to inability to remove systematics as well as the inaccuracy of molecular line lists.
Despite this explaination, \cite{Pelletier2023} was able to detect VO with both ESPRESSO and MAROON-X, suggesting that in the optical regime, molecular line lists can be accurate enough to detect VO.
Discrepancies have also occurred for molecules observable in the infrared; for instance, \cite{Lockwood2014} reported a detection of H$_2$O absorption in $\tau$ Boo b using $L$-band NIRSPEC spectra, but \cite{Pelletier2021} did not detect H$_2$O with $YHJK$ spectra from SPIRou.
The variation between results for not only molecular species, but the atomic species in other recent works discussed in this paper, accentuates the need to reevaluate the current methods frequently used to assess the the significance of detections using high resolution spectroscopy.

Beyond the problem of inaccurate line lists, methodological issues may be responsible for some of these discrepancies. \cite{Cabot2019} highlighted that spurious detections can be caused by incorrect optimization of detrending. It is also well-known that the detection of the minute signals due to exoplanet atmospheres hinges critically upon the removal of telluric and stellar signals and instrumental systematics, and there is not yet a consensus upon the best methodology \citep[e.g.,][]{Johnson2022}. 
Other recent works have suggested using a higher ($5\sigma$) threshold for detections \citep[e.g.,][]{Borsato2023}. The use of Bayesian log-likelihood methodology \citep[e.g.,][]{BrogiLine2017,Gibson2020} may also allow for more robust conclusions on the significance of a given signal. While high-resolution spectroscopic techniques have advanced greatly over the nearly two and a half decades since the initial efforts of \cite{Collier99} and \cite{Charbonneau99}, clearly there is still more work to be done to understand and optimize our methodology to mitigate the effects of systematics and detect planetary signals. 

One impediment to this advancement is that the underlying data are frequently not publicly available, preventing subsequent analysis with different methods to verify results and test methodologies. As a concrete example, none of the previous high-resolution emission datasets on KELT-20b--the HARPS-N data used by \cite{Borsa2022}, the  data used by \cite{Kasper2022}, or the CARMENES data used by \cite{Cont2021a}--are publicly available, to the best of our knowledge. Public availability of data aids in the assessment of the methodological robustness of data analysis, and without access to the data underlying previous publications it is impossible for us definitively determine the reasons for discrepancies in our results. In order to allow for greater reproducability and future comparisons, we are releasing the PEPSI data that we used in this paper through the NASA Exoplanet Archive.

We are unable to present a definitive explanation for our non-detections, however these results leave us with multiple avenues to further investigate including more detailed chemical modeling of KELT-20b's atmosphere and low-resolution observations to search for sources of continuum opacity.
With these avenues fully explored in the future, we hope to fully constrain the chemical constituents that make up the atmosphere of KELT-20b, allowing us to gain insight into its atmospheric dynamics and evolutionary history.

\section*{Acknowledgements}

We would like to thank Paul Molliere and Aurora Kesseli for their help in converting the DACE opacities into a usable form for \texttt{petitRADTRANS}.

S.P would like to thank M.C.J., as well and the rest of Ohio State's Summer Undergraduate Research Program, for providing the opportunity to take on this project.

M.C.J. was supported in part by NASA Grant 80NSSC23K0692.
B.S.G. was supported by the Thomas Jefferson Chair for Space Exploration endowment from the Ohio State University.
K.P. acknowledges funding
from the German \textit{Leibniz Community} under project number P67/2018.


The LBT is an international collaboration among institutions in the United States, Italy, and Germany. LBT Corporation Members are: The Ohio State University, representing OSU, University of Notre Dame, University of Minnesota, and University of Virginia; LBT Beteiligungsgesellschaft, Germany, representing the Max-Planck Society, The Leibniz Institute for Astrophysics Potsdam, and Heidelberg University; The University of Arizona on behalf of the Arizona Board of Regents; and the Istituto Nazionale di Astrofisica, Italy.

\section*{Data Availability}

The PEPSI data underlying this paper will be made publicly available at the NASA Exoplanet Archive.


\bibliographystyle{mnras}
\bibliography{example} 

\begin{thebibliography}{}
\makeatletter
\relax
\def\mn@urlcharsother{\let\do\@makeother \do\$\do\&\do\#\do\^\do\_\do\%\do\~}
\def\mn@doi{\begingroup\mn@urlcharsother \@ifnextchar [ {\mn@doi@}
  {\mn@doi@[]}}
\def\mn@doi@[#1]#2{\def\@tempa{#1}\ifx\@tempa\@empty \href
  {http://dx.doi.org/#2} {doi:#2}\else \href {http://dx.doi.org/#2} {#1}\fi
  \endgroup}
\def\mn@eprint#1#2{\mn@eprint@#1:#2::\@nil}
\def\mn@eprint@arXiv#1{\href {http://arxiv.org/abs/#1} {{\tt arXiv:#1}}}
\def\mn@eprint@dblp#1{\href {http://dblp.uni-trier.de/rec/bibtex/#1.xml}
  {dblp:#1}}
\def\mn@eprint@#1:#2:#3:#4\@nil{\def\@tempa {#1}\def\@tempb {#2}\def\@tempc
  {#3}\ifx \@tempc \@empty \let \@tempc \@tempb \let \@tempb \@tempa \fi \ifx
  \@tempb \@empty \def\@tempb {arXiv}\fi \@ifundefined
  {mn@eprint@\@tempb}{\@tempb:\@tempc}{\expandafter \expandafter \csname
  mn@eprint@\@tempb\endcsname \expandafter{\@tempc}}}

\bibitem[\protect\citeauthoryear{{Arcangeli} et~al.,}{{Arcangeli}
  et~al.}{2018}]{Arcangeli2018}
{Arcangeli} J.,  et~al., 2018, \mn@doi [\apjl] {10.3847/2041-8213/aab272},
  \href {https://ui.adsabs.harvard.edu/abs/2018ApJ...855L..30A} {855, L30}

\bibitem[\protect\citeauthoryear{{Bello-Arufe}, {Buchhave}, {Mendon{\c{c}}a},
  {Tronsgaard}, {Heng}, {Jens Hoeijmakers}  \& {Mayo}}{{Bello-Arufe}
  et~al.}{2022}]{BelloArufe2022}
{Bello-Arufe} A.,  {Buchhave} L.~A.,  {Mendon{\c{c}}a} J.~M.,  {Tronsgaard} R.,
   {Heng} K.,  {Jens Hoeijmakers} H.,   {Mayo} A.~W.,  2022, \mn@doi [\aap]
  {10.1051/0004-6361/202142787}, \href
  {https://ui.adsabs.harvard.edu/abs/2022A&A...662A..51B} {662, A51}

\bibitem[\protect\citeauthoryear{{Beltz}, {Rauscher}, {Kempton}, {Malsky},
  {Ochs}, {Arora}  \& {Savel}}{{Beltz} et~al.}{2022}]{Beltz}
{Beltz} H.,  {Rauscher} E.,  {Kempton} E. M.~R.,  {Malsky} I.,  {Ochs} G.,
  {Arora} M.,   {Savel} A.,  2022, \mn@doi [\aj] {10.3847/1538-3881/ac897b},
  \href {https://ui.adsabs.harvard.edu/abs/2022AJ....164..140B} {164, 140}

\bibitem[\protect\citeauthoryear{{Borsa} et~al.,}{{Borsa}
  et~al.}{2022}]{Borsa2022}
{Borsa} F.,  et~al., 2022, \mn@doi [\aap] {10.1051/0004-6361/202142768}, \href
  {https://ui.adsabs.harvard.edu/abs/2022A&A...663A.141B} {663, A141}

\bibitem[\protect\citeauthoryear{{Borsato}, {Hoeijmakers}, {Prinoth},
  {Thorsbro}, {Forsberg}, {Kitzmann}, {Jones}  \& {Heng}}{{Borsato}
  et~al.}{2023}]{Borsato2023}
{Borsato} N.~W.,  {Hoeijmakers} H.~J.,  {Prinoth} B.,  {Thorsbro} B.,
  {Forsberg} R.,  {Kitzmann} D.,  {Jones} K.,   {Heng} K.,  2023, \mn@doi
  [\aap] {10.1051/0004-6361/202245121}, \href
  {https://ui.adsabs.harvard.edu/abs/2023A&A...673A.158B} {673, A158}

\bibitem[\protect\citeauthoryear{{Brogi}, {Snellen}, {de Kok}, {Albrecht},
  {Birkby}  \& {de Mooij}}{{Brogi} et~al.}{2012}]{Brogi2012}
{Brogi} M.,  {Snellen} I. A.~G.,  {de Kok} R.~J.,  {Albrecht} S.,  {Birkby} J.,
    {de Mooij} E. J.~W.,  2012, \mn@doi [\nat] {10.1038/nature11161}, \href
  {https://ui.adsabs.harvard.edu/abs/2012Natur.486..502B} {486, 502}

\bibitem[\protect\citeauthoryear{{Brogi}, {Line}, {Bean}, {D{\'e}sert}  \&
  {Schwarz}}{{Brogi} et~al.}{2017}]{BrogiLine2017}
{Brogi} M.,  {Line} M.,  {Bean} J.,  {D{\'e}sert} J.~M.,   {Schwarz} H.,  2017,
  \mn@doi [\apjl] {10.3847/2041-8213/aa6933}, \href
  {https://ui.adsabs.harvard.edu/abs/2017ApJ...839L...2B} {839, L2}

\bibitem[\protect\citeauthoryear{{Brogi} et~al.,}{{Brogi}
  et~al.}{2023}]{Brogi2023}
{Brogi} M.,  et~al., 2023, \mn@doi [\aj] {10.3847/1538-3881/acaf5c}, \href
  {https://ui.adsabs.harvard.edu/abs/2023AJ....165...91B} {165, 91}

\bibitem[\protect\citeauthoryear{{Cabot}, {Madhusudhan}, {Hawker}  \&
  {Gandhi}}{{Cabot} et~al.}{2019}]{Cabot2019}
{Cabot} S. H.~C.,  {Madhusudhan} N.,  {Hawker} G.~A.,   {Gandhi} S.,  2019,
  \mn@doi [\mnras] {10.1093/mnras/sty2994}, \href
  {https://ui.adsabs.harvard.edu/abs/2019MNRAS.482.4422C} {482, 4422}

\bibitem[\protect\citeauthoryear{{Casasayas-Barris} et~al.,}{{Casasayas-Barris}
  et~al.}{2019}]{Casasayas2019}
{Casasayas-Barris} N.,  et~al., 2019, \mn@doi [\aap]
  {10.1051/0004-6361/201935623}, \href
  {https://ui.adsabs.harvard.edu/abs/2019A&A...628A...9C} {628, A9}

\bibitem[\protect\citeauthoryear{{Charbonneau}, {Noyes}, {Korzennik},
  {Nisenson}, {Jha}, {Vogt}  \& {Kibrick}}{{Charbonneau}
  et~al.}{1999}]{Charbonneau99}
{Charbonneau} D.,  {Noyes} R.~W.,  {Korzennik} S.~G.,  {Nisenson} P.,  {Jha}
  S.,  {Vogt} S.~S.,   {Kibrick} R.~I.,  1999, \mn@doi [\apjl]
  {10.1086/312234}, \href
  {https://ui.adsabs.harvard.edu/abs/1999ApJ...522L.145C} {522, L145}

\bibitem[\protect\citeauthoryear{{Charbonneau}, {Brown}, {Noyes}  \&
  {Gilliland}}{{Charbonneau} et~al.}{2002}]{Charbonneau2002}
{Charbonneau} D.,  {Brown} T.~M.,  {Noyes} R.~W.,   {Gilliland} R.~L.,  2002,
  \mn@doi [\apj] {10.1086/338770}, \href
  {https://ui.adsabs.harvard.edu/abs/2002ApJ...568..377C} {568, 377}

\bibitem[\protect\citeauthoryear{{Collier Cameron}, {Horne}, {Penny}  \&
  {James}}{{Collier Cameron} et~al.}{1999}]{Collier99}
{Collier Cameron} A.,  {Horne} K.,  {Penny} A.,   {James} D.,  1999, \mn@doi
  [\nat] {10.1038/45451}, \href
  {https://ui.adsabs.harvard.edu/abs/1999Natur.402..751C} {402, 751}

\bibitem[\protect\citeauthoryear{{Collier Cameron}, {Horne}, {James}, {Penny}
  \& {Semel}}{{Collier Cameron} et~al.}{2000}]{Collier2000}
{Collier Cameron} A.,  {Horne} K.,  {James} D.,  {Penny} A.,   {Semel} M.,
  2000, \mn@doi [arXiv e-prints] {10.48550/arXiv.astro-ph/0012186}, \href
  {https://ui.adsabs.harvard.edu/abs/2000astro.ph.12186C} {pp
  astro--ph/0012186}

\bibitem[\protect\citeauthoryear{{Cont} et~al.,}{{Cont}
  et~al.}{2021}]{Cont2021b}
{Cont} D.,  et~al., 2021, \mn@doi [\aap] {10.1051/0004-6361/202140732}, \href
  {https://ui.adsabs.harvard.edu/abs/2021A&A...651A..33C} {651, A33}

\bibitem[\protect\citeauthoryear{{Cont} et~al.,}{{Cont}
  et~al.}{2022}]{Cont2021a}
{Cont} D.,  et~al., 2022, \mn@doi [\aap] {10.1051/0004-6361/202142776}, \href
  {https://ui.adsabs.harvard.edu/abs/2022A&A...657L...2C} {657, L2}

\bibitem[\protect\citeauthoryear{{Evans} et~al.,}{{Evans}
  et~al.}{2017}]{Evans2017}
{Evans} T.~M.,  et~al., 2017, \mn@doi [\nat] {10.1038/nature23266}, \href
  {https://ui.adsabs.harvard.edu/abs/2017Natur.548...58E} {548, 58}

\bibitem[\protect\citeauthoryear{{Fu} et~al.,}{{Fu} et~al.}{2022}]{Fu2022}
{Fu} G.,  et~al., 2022, \mn@doi [\apjl] {10.3847/2041-8213/ac4968}, \href
  {https://ui.adsabs.harvard.edu/abs/2022ApJ...925L...3F} {925, L3}

\bibitem[\protect\citeauthoryear{Gandhi \& Madhusudhan}{Gandhi \&
  Madhusudhan}{2019}]{Gandhi2019}
Gandhi S.,  Madhusudhan N.,  2019, \mn@doi [Monthly Notices of the Royal
  Astronomical Society] {10.1093/mnras/stz751}, 485, 5817

\bibitem[\protect\citeauthoryear{{Gandhi} et~al.,}{{Gandhi}
  et~al.}{2023}]{Gandhi2023}
{Gandhi} S.,  et~al., 2023, \mn@doi [\aj] {10.3847/1538-3881/accd65}, \href
  {https://ui.adsabs.harvard.edu/abs/2023AJ....165..242G} {165, 242}

\bibitem[\protect\citeauthoryear{{Gibson} et~al.,}{{Gibson}
  et~al.}{2020}]{Gibson2020}
{Gibson} N.~P.,  et~al., 2020, \mn@doi [\mnras] {10.1093/mnras/staa228}, \href
  {https://ui.adsabs.harvard.edu/abs/2020MNRAS.493.2215G} {493, 2215}

\bibitem[\protect\citeauthoryear{{Gray}}{{Gray}}{2005}]{Gray}
{Gray} D.~F.,  2005, The Observation and Analysis of Stellar Photospheres, 3rd
  edition edn.
Cambridge University Press, Cambridge

\bibitem[\protect\citeauthoryear{{Guillot}}{{Guillot}}{2010}]{Guillot}
{Guillot} T.,  2010, \mn@doi [\aap] {10.1051/0004-6361/200913396}, \href
  {https://ui.adsabs.harvard.edu/abs/2010A&A...520A..27G} {520, A27}

\bibitem[\protect\citeauthoryear{{Herman}, {de Mooij}, {Jayawardhana}  \&
  {Brogi}}{{Herman} et~al.}{2020}]{Herman2020}
{Herman} M.~K.,  {de Mooij} E. J.~W.,  {Jayawardhana} R.,   {Brogi} M.,  2020,
  \mn@doi [\aj] {10.3847/1538-3881/ab9e77}, \href
  {https://ui.adsabs.harvard.edu/abs/2020AJ....160...93H} {160, 93}

\bibitem[\protect\citeauthoryear{{Herman}, {de Mooij}, {Nugroho}, {Gibson}  \&
  {Jayawardhana}}{{Herman} et~al.}{2022}]{Herman2022}
{Herman} M.~K.,  {de Mooij} E. J.~W.,  {Nugroho} S.~K.,  {Gibson} N.~P.,
  {Jayawardhana} R.,  2022, \mn@doi [\aj] {10.3847/1538-3881/ac5f4d}, \href
  {https://ui.adsabs.harvard.edu/abs/2022AJ....163..248H} {163, 248}

\bibitem[\protect\citeauthoryear{{Hoeijmakers} et~al.,}{{Hoeijmakers}
  et~al.}{2020}]{Hoeijmakers2020}
{Hoeijmakers} H.~J.,  et~al., 2020, \mn@doi [A\&A]
  {10.1051/0004-6361/202038365}, 641, A123

\bibitem[\protect\citeauthoryear{{JWST Transiting Exoplanet Community Early
  Release Science Team} et~al.,}{{JWST Transiting Exoplanet Community Early
  Release Science Team} et~al.}{2023}]{JWST}
{JWST Transiting Exoplanet Community Early Release Science Team} et~al., 2023,
  \mn@doi [\nat] {10.1038/s41586-022-05269-w}, \href
  {https://ui.adsabs.harvard.edu/abs/2023Natur.614..649J} {614, 649}

\bibitem[\protect\citeauthoryear{{Johnson} et~al.,}{{Johnson}
  et~al.}{2023}]{Johnson2022}
{Johnson} M.~C.,  et~al., 2023, \mn@doi [\aj] {10.3847/1538-3881/acb7e2}, \href
  {https://ui.adsabs.harvard.edu/abs/2023AJ....165..157J} {165, 157}

\bibitem[\protect\citeauthoryear{{Kasper}, {Bean}, {Line}, {Seifahrt},
  {St{\"u}rmer}, {Pino}, {D{\'e}sert}  \& {Brogi}}{{Kasper}
  et~al.}{2021}]{Kasper2021}
{Kasper} D.,  {Bean} J.~L.,  {Line} M.~R.,  {Seifahrt} A.,  {St{\"u}rmer} J.,
  {Pino} L.,  {D{\'e}sert} J.-M.,   {Brogi} M.,  2021, \mn@doi [\apjl]
  {10.3847/2041-8213/ac30e1}, \href
  {https://ui.adsabs.harvard.edu/abs/2021ApJ...921L..18K} {921, L18}

\bibitem[\protect\citeauthoryear{{Kasper} et~al.,}{{Kasper}
  et~al.}{2023}]{Kasper2022}
{Kasper} D.,  et~al., 2023, \mn@doi [\aj] {10.3847/1538-3881/ac9f40}, \href
  {https://ui.adsabs.harvard.edu/abs/2023AJ....165....7K} {165, 7}

\bibitem[\protect\citeauthoryear{{Kausch} et~al.,}{{Kausch}
  et~al.}{2015}]{Kausch2015}
{Kausch} W.,  et~al., 2015, \mn@doi [\aap] {10.1051/0004-6361/201423909}, \href
  {https://ui.adsabs.harvard.edu/abs/2015A&A...576A..78K} {576, A78}

\bibitem[\protect\citeauthoryear{{Keles} et~al.,}{{Keles}
  et~al.}{2022}]{Keles2022}
{Keles} E.,  et~al., 2022, \mn@doi [\mnras] {10.1093/mnras/stac810}, \href
  {https://ui.adsabs.harvard.edu/abs/2022MNRAS.513.1544K} {513, 1544}

\bibitem[\protect\citeauthoryear{Kesseli, Snellen, Alonso-Floriano, Mollière
  \& Serindag}{Kesseli et~al.}{2020}]{Kesseli2020}
Kesseli A.~Y.,  Snellen I. A.~G.,  Alonso-Floriano F.~J.,  Mollière P.,
  Serindag D.~B.,  2020, \mn@doi [The Astronomical Journal]
  {10.3847/1538-3881/abb59c}, 160, 228

\bibitem[\protect\citeauthoryear{{Kesseli}, {Snellen}, {Casasayas-Barris},
  {Molli{\`e}re}  \& {S{\'a}nchez-L{\'o}pez}}{{Kesseli}
  et~al.}{2022}]{Kesseli2022}
{Kesseli} A.~Y.,  {Snellen} I.~A.~G.,  {Casasayas-Barris} N.,  {Molli{\`e}re}
  P.,   {S{\'a}nchez-L{\'o}pez} A.,  2022, \mn@doi [\aj]
  {10.3847/1538-3881/ac4336}, \href
  {https://ui.adsabs.harvard.edu/abs/2022AJ....163..107K} {163, 107}

\bibitem[\protect\citeauthoryear{{Kochukhov}, {Makaganiuk}  \&
  {Piskunov}}{{Kochukhov} et~al.}{2010}]{Kochukhov2010}
{Kochukhov} O.,  {Makaganiuk} V.,   {Piskunov} N.,  2010, \mn@doi [\aap]
  {10.1051/0004-6361/201015429}, \href
  {https://ui.adsabs.harvard.edu/abs/2010A&A...524A...5K} {524, A5}

\bibitem[\protect\citeauthoryear{{Leigh}, {Collier Cameron}, {Horne}, {Penny}
  \& {James}}{{Leigh} et~al.}{2003}]{Leigh2003}
{Leigh} C.,  {Collier Cameron} A.,  {Horne} K.,  {Penny} A.,   {James} D.,
  2003, \mn@doi [\mnras] {10.1046/j.1365-8711.2003.06901.x}, \href
  {https://ui.adsabs.harvard.edu/abs/2003MNRAS.344.1271L} {344, 1271}

\bibitem[\protect\citeauthoryear{{Line} et~al.,}{{Line}
  et~al.}{2021}]{Line2021}
{Line} M.~R.,  et~al., 2021, \mn@doi [\nat] {10.1038/s41586-021-03912-6}, \href
  {https://ui.adsabs.harvard.edu/abs/2021Natur.598..580L} {598, 580}

\bibitem[\protect\citeauthoryear{{Lockwood}, {Johnson}, {Bender}, {Carr},
  {Barman}, {Richert}  \& {Blake}}{{Lockwood} et~al.}{2014}]{Lockwood2014}
{Lockwood} A.~C.,  {Johnson} J.~A.,  {Bender} C.~F.,  {Carr} J.~S.,  {Barman}
  T.,  {Richert} A. J.~W.,   {Blake} G.~A.,  2014, \mn@doi [\apjl]
  {10.1088/2041-8205/783/2/L29}, \href
  {https://ui.adsabs.harvard.edu/abs/2014ApJ...783L..29L} {783, L29}

\bibitem[\protect\citeauthoryear{{Lothringer}, {Barman}  \&
  {Koskinen}}{{Lothringer} et~al.}{2018}]{Lothringer2018}
{Lothringer} J.~D.,  {Barman} T.,   {Koskinen} T.,  2018, \mn@doi [\apj]
  {10.3847/1538-4357/aadd9e}, \href
  {https://ui.adsabs.harvard.edu/abs/2018ApJ...866...27L} {866, 27}

\bibitem[\protect\citeauthoryear{Lund et~al.,}{Lund et~al.}{2017}]{Lund2017}
Lund M.~B.,  et~al., 2017, \mn@doi [The Astronomical Journal]
  {10.3847/1538-3881/aa8f95}, 154, 194

\bibitem[\protect\citeauthoryear{{Merritt} et~al.,}{{Merritt}
  et~al.}{2020}]{Merritt2020}
{Merritt} S.~R.,  et~al., 2020, \mn@doi [\aap] {10.1051/0004-6361/201937409},
  \href {https://ui.adsabs.harvard.edu/abs/2020A&A...636A.117M} {636, A117}

\bibitem[\protect\citeauthoryear{{Molli{\`e}re}, {Wardenier}, {van Boekel},
  {Henning}, {Molaverdikhani}  \& {Snellen}}{{Molli{\`e}re}
  et~al.}{2019}]{Molliere2019}
{Molli{\`e}re} P.,  {Wardenier} J.~P.,  {van Boekel} R.,  {Henning} T.,
  {Molaverdikhani} K.,   {Snellen} I.~A.~G.,  2019, \mn@doi [\aap]
  {10.1051/0004-6361/201935470}, \href
  {https://ui.adsabs.harvard.edu/abs/2019A&A...627A..67M} {627, A67}

\bibitem[\protect\citeauthoryear{{Nugroho}, {Kawahara}, {Masuda}, {Hirano},
  {Kotani}  \& {Tajitsu}}{{Nugroho} et~al.}{2017}]{Nugroho2017}
{Nugroho} S.~K.,  {Kawahara} H.,  {Masuda} K.,  {Hirano} T.,  {Kotani} T.,
  {Tajitsu} A.,  2017, \mn@doi [\aj] {10.3847/1538-3881/aa9433}, \href
  {https://ui.adsabs.harvard.edu/abs/2017AJ....154..221N} {154, 221}

\bibitem[\protect\citeauthoryear{{Nugroho}, {Gibson}, {de Mooij}, {Watson},
  {Kawahara}  \& {Merritt}}{{Nugroho} et~al.}{2020}]{Nugroho2020}
{Nugroho} S.~K.,  {Gibson} N.~P.,  {de Mooij} E. J.~W.,  {Watson} C.~A.,
  {Kawahara} H.,   {Merritt} S.,  2020, \mn@doi [\mnras]
  {10.1093/mnras/staa1459}, \href
  {https://ui.adsabs.harvard.edu/abs/2020MNRAS.496..504N} {496, 504}

\bibitem[\protect\citeauthoryear{Ouyang et~al.,}{Ouyang
  et~al.}{2023}]{Qinglin2023}
Ouyang Q.,  et~al., 2023, Detection of TiO and VO in the atmosphere of
  WASP-121b and Evidence for its temporal variation (\mn@eprint {arXiv}
  {2304.00461})

\bibitem[\protect\citeauthoryear{{Pai Asnodkar}, {Wang}, {Gaudi}, {Cauley},
  {Eastman}, {Ilyin}, {Strassmeier}  \& {Beatty}}{{Pai Asnodkar}
  et~al.}{2022a}]{Pai2022b}
{Pai Asnodkar} A.,  {Wang} J.,  {Gaudi} B.~S.,  {Cauley} P.~W.,  {Eastman}
  J.~D.,  {Ilyin} I.,  {Strassmeier} K.,   {Beatty} T.,  2022a, \mn@doi [\aj]
  {10.3847/1538-3881/ac32c7}, \href
  {https://ui.adsabs.harvard.edu/abs/2022AJ....163...40P} {163, 40}

\bibitem[\protect\citeauthoryear{{Pai Asnodkar}, {Wang}, {Eastman}, {Cauley},
  {Gaudi}, {Ilyin}  \& {Strassmeier}}{{Pai Asnodkar} et~al.}{2022b}]{Pai2022}
{Pai Asnodkar} A.,  {Wang} J.,  {Eastman} J.~D.,  {Cauley} P.~W.,  {Gaudi}
  B.~S.,  {Ilyin} I.,   {Strassmeier} K.,  2022b, \mn@doi [\aj]
  {10.3847/1538-3881/ac51d2}, \href
  {https://ui.adsabs.harvard.edu/abs/2022AJ....163..155P} {163, 155}

\bibitem[\protect\citeauthoryear{{Parmentier}, {Showman}  \&
  {Lian}}{{Parmentier} et~al.}{2013}]{Parmentier2013}
{Parmentier} V.,  {Showman} A.~P.,   {Lian} Y.,  2013, \mn@doi [\aap]
  {10.1051/0004-6361/201321132}, \href
  {https://ui.adsabs.harvard.edu/abs/2013A&A...558A..91P} {558, A91}

\bibitem[\protect\citeauthoryear{{Parviainen}}{{Parviainen}}{2023}]{Parviainen2023}
{Parviainen} H.,  2023, \mn@doi [\aap] {10.1051/0004-6361/202345937}, \href
  {https://ui.adsabs.harvard.edu/abs/2023A&A...671L...3P} {671, L3}

\bibitem[\protect\citeauthoryear{{Pelletier} et~al.,}{{Pelletier}
  et~al.}{2021}]{Pelletier2021}
{Pelletier} S.,  et~al., 2021, \mn@doi [\aj] {10.3847/1538-3881/ac0428}, \href
  {https://ui.adsabs.harvard.edu/abs/2021AJ....162...73P} {162, 73}

\bibitem[\protect\citeauthoryear{{Pelletier} et~al.,}{{Pelletier}
  et~al.}{2023}]{Pelletier2023}
{Pelletier} S.,  et~al., 2023, \mn@doi [\nat] {10.1038/s41586-023-06134-0},
  \href {https://ui.adsabs.harvard.edu/abs/2023Natur.619..491P} {619, 491}

\bibitem[\protect\citeauthoryear{{Prinoth} et~al.,}{{Prinoth}
  et~al.}{2022}]{Prinoth2022}
{Prinoth} B.,  et~al., 2022, \mn@doi [Nature Astronomy]
  {10.1038/s41550-021-01581-z}, \href
  {https://ui.adsabs.harvard.edu/abs/2022NatAs...6..449P} {6, 449}

\bibitem[\protect\citeauthoryear{Rivlin, Lodi, Yurchenko, Tennyson  \&
  Le~Roy}{Rivlin et~al.}{2015}]{Rivlin2015}
Rivlin T.,  Lodi L.,  Yurchenko S.~N.,  Tennyson J.,   Le~Roy R.~J.,  2015,
  \mn@doi [Monthly Notices of the Royal Astronomical Society]
  {10.1093/mnras/stv979}, 451, 634

\bibitem[\protect\citeauthoryear{{Rodler}, {K{\"u}rster}  \&
  {Henning}}{{Rodler} et~al.}{2010}]{Rodler2010}
{Rodler} F.,  {K{\"u}rster} M.,   {Henning} T.,  2010, \mn@doi [\aap]
  {10.1051/0004-6361/200913627}, \href
  {https://ui.adsabs.harvard.edu/abs/2010A&A...514A..23R} {514, A23}

\bibitem[\protect\citeauthoryear{{Rodler}, {K{\"u}rster}, {L{\'o}pez-Morales}
  \& {Ribas}}{{Rodler} et~al.}{2013}]{Rodler2013}
{Rodler} F.,  {K{\"u}rster} M.,  {L{\'o}pez-Morales} M.,   {Ribas} I.,  2013,
  \mn@doi [Astronomische Nachrichten] {10.1002/asna.201211744}, \href
  {https://ui.adsabs.harvard.edu/abs/2013AN....334..188R} {334, 188}

\bibitem[\protect\citeauthoryear{{Scandariato} et~al.,}{{Scandariato}
  et~al.}{2021}]{Scandariato2021}
{Scandariato} G.,  et~al., 2021, \mn@doi [\aap] {10.1051/0004-6361/202039271},
  \href {https://ui.adsabs.harvard.edu/abs/2021A&A...646A.159S} {646, A159}

\bibitem[\protect\citeauthoryear{{Scandariato} et~al.,}{{Scandariato}
  et~al.}{2023}]{Scandariato2023}
{Scandariato} G.,  et~al., 2023, \mn@doi [arXiv e-prints]
  {10.48550/arXiv.2304.03328}, \href
  {https://ui.adsabs.harvard.edu/abs/2023arXiv230403328S} {p. arXiv:2304.03328}

\bibitem[\protect\citeauthoryear{{Serindag}, {Nugroho}, {Molli\`ere}, {de
  Mooij}, {Gibson}  \& {Snellen}}{{Serindag} et~al.}{2021}]{Serindag2021}
{Serindag} D.~B.,  {Nugroho} S.~K.,  {Molli\`ere} P.,  {de Mooij} E.~J.~W.,
  {Gibson} N.~P.,   {Snellen} I.~A.~G.,  2021, \mn@doi [A\&A]
  {10.1051/0004-6361/202039135}, 645, A90

\bibitem[\protect\citeauthoryear{{Smette} et~al.,}{{Smette}
  et~al.}{2015}]{Smette2015}
{Smette} A.,  et~al., 2015, \mn@doi [\aap] {10.1051/0004-6361/201423932}, \href
  {https://ui.adsabs.harvard.edu/abs/2015A&A...576A..77S} {576, A77}

\bibitem[\protect\citeauthoryear{{Stangret}, {Casasayas-Barris}, {Pall{\'e}},
  {Yan}, {S{\'a}nchez-L{\'o}pez}  \& {L{\'o}pez-Puertas}}{{Stangret}
  et~al.}{2020}]{Stangret2020}
{Stangret} M.,  {Casasayas-Barris} N.,  {Pall{\'e}} E.,  {Yan} F.,
  {S{\'a}nchez-L{\'o}pez} A.,   {L{\'o}pez-Puertas} M.,  2020, \mn@doi [\aap]
  {10.1051/0004-6361/202037541}, \href
  {https://ui.adsabs.harvard.edu/abs/2020A&A...638A..26S} {638, A26}

\bibitem[\protect\citeauthoryear{{Stock}, {Kitzmann}, {Patzer}  \&
  {Sedlmayr}}{{Stock} et~al.}{2018}]{FastChem}
{Stock} J.~W.,  {Kitzmann} D.,  {Patzer} A. B.~C.,   {Sedlmayr} E.,  2018,
  \mn@doi [\mnras] {10.1093/mnras/sty1531}, \href
  {https://ui.adsabs.harvard.edu/abs/2018MNRAS.479..865S} {479, 865}

\bibitem[\protect\citeauthoryear{{Strassmeier} et~al.,}{{Strassmeier}
  et~al.}{2015}]{Strassmeier2015}
{Strassmeier} K.~G.,  et~al., 2015, \mn@doi [Astronomische Nachrichten]
  {10.1002/asna.201512172}, \href
  {https://ui.adsabs.harvard.edu/abs/2015AN....336..324S} {336, 324}

\bibitem[\protect\citeauthoryear{{Talens} et~al.,}{{Talens}
  et~al.}{2018}]{Talens2018}
{Talens} G. J.~J.,  et~al., 2018, \mn@doi [A\&A] {10.1051/0004-6361/201731512},
  612, A57

\bibitem[\protect\citeauthoryear{Tamuz, Mazeh  \& Zucker}{Tamuz
  et~al.}{2005}]{Tamuz2005}
Tamuz O.,  Mazeh T.,   Zucker S.,  2005, \mn@doi [Monthly Notices of the Royal
  Astronomical Society] {10.1111/j.1365-2966.2004.08585.x}, 356, 1466

\bibitem[\protect\citeauthoryear{{Valenti} \& {Piskunov}}{{Valenti} \&
  {Piskunov}}{1996}]{ValentiPiskunov1996}
{Valenti} J.~A.,  {Piskunov} N.,  1996, \aaps, \href
  {https://ui.adsabs.harvard.edu/abs/1996A&AS..118..595V} {118, 595}

\bibitem[\protect\citeauthoryear{{Valenti} \& {Piskunov}}{{Valenti} \&
  {Piskunov}}{2012}]{ValentiPiskunov2012}
{Valenti} J.~A.,  {Piskunov} N.,  2012, {SME: Spectroscopy Made Easy},
  Astrophysics Source Code Library, record ascl:1202.013 (\mn@eprint {ascl}
  {1202.013})

\bibitem[\protect\citeauthoryear{{Visscher}, {Lodders}  \& {Fegley}}{{Visscher}
  et~al.}{2010}]{Visscher2010}
{Visscher} C.,  {Lodders} K.,   {Fegley} Bruce J.,  2010, \mn@doi [\apj]
  {10.1088/0004-637X/716/2/1060}, \href
  {https://ui.adsabs.harvard.edu/abs/2010ApJ...716.1060V} {716, 1060}

\bibitem[\protect\citeauthoryear{{Wang}, {Prato}  \& {Mawet}}{{Wang}
  et~al.}{2017}]{Wang2017}
{Wang} J.,  {Prato} L.,   {Mawet} D.,  2017, \mn@doi [\apj]
  {10.3847/1538-4357/aa6345}, \href
  {https://ui.adsabs.harvard.edu/abs/2017ApJ...838...35W} {838, 35}

\bibitem[\protect\citeauthoryear{Yadin, Veness, Conti, Hill, Yurchenko  \&
  Tennyson}{Yadin et~al.}{2012}]{Yadin2012}
Yadin B.,  Veness T.,  Conti P.,  Hill C.,  Yurchenko S.~N.,   Tennyson J.,
  2012, arXiv: Solar and Stellar Astrophysics

\bibitem[\protect\citeauthoryear{{Yan} et~al.,}{{Yan} et~al.}{2020}]{Yan2020}
{Yan} F.,  et~al., 2020, \mn@doi [A\&A] {10.1051/0004-6361/202038294}, 640, L5

\bibitem[\protect\citeauthoryear{{Yan} et~al.,}{{Yan} et~al.}{2022a}]{Yan2022b}
{Yan} F.,  et~al., 2022a, \mn@doi [\aap] {10.1051/0004-6361/202142395}, \href
  {https://ui.adsabs.harvard.edu/abs/2022A&A...659A...7Y} {659, A7}

\bibitem[\protect\citeauthoryear{{Yan} et~al.,}{{Yan} et~al.}{2022b}]{Yan2022a}
{Yan} F.,  et~al., 2022b, \mn@doi [A\&A] {10.1051/0004-6361/202243503}, 661, L6

\bibitem[\protect\citeauthoryear{{Yan} et~al.,}{{Yan} et~al.}{2023}]{Yan2023}
{Yan} F.,  et~al., 2023, \mn@doi [\aap] {10.1051/0004-6361/202245371}, \href
  {https://ui.adsabs.harvard.edu/abs/2023A&A...672A.107Y} {672, A107}

\bibitem[\protect\citeauthoryear{{Zilinskas}, {van Buchem}, {Miguel}, {Louca},
  {Lupu}, {Zieba}  \& {van Westrenen}}{{Zilinskas}
  et~al.}{2022}]{Zilinskas2022}
{Zilinskas} M.,  {van Buchem} C.~P.~A.,  {Miguel} Y.,  {Louca} A.,  {Lupu} R.,
  {Zieba} S.,   {van Westrenen} W.,  2022, \mn@doi [\aap]
  {10.1051/0004-6361/202142984}, \href
  {https://ui.adsabs.harvard.edu/abs/2022A&A...661A.126Z} {661, A126}

\bibitem[\protect\citeauthoryear{{van Sluijs} et~al.,}{{van Sluijs}
  et~al.}{2022}]{vanSlujis2022}
{van Sluijs} L.,  et~al., 2022, \mn@doi [arXiv e-prints]
  {10.48550/arXiv.2203.13234}, \href
  {https://ui.adsabs.harvard.edu/abs/2022arXiv220313234V} {p. arXiv:2203.13234}

\makeatother
\end{thebibliography}


\appendix

\section{Molecular Non-Detections}

\begin{figure*}
    	\includegraphics[width=0.4\textwidth]{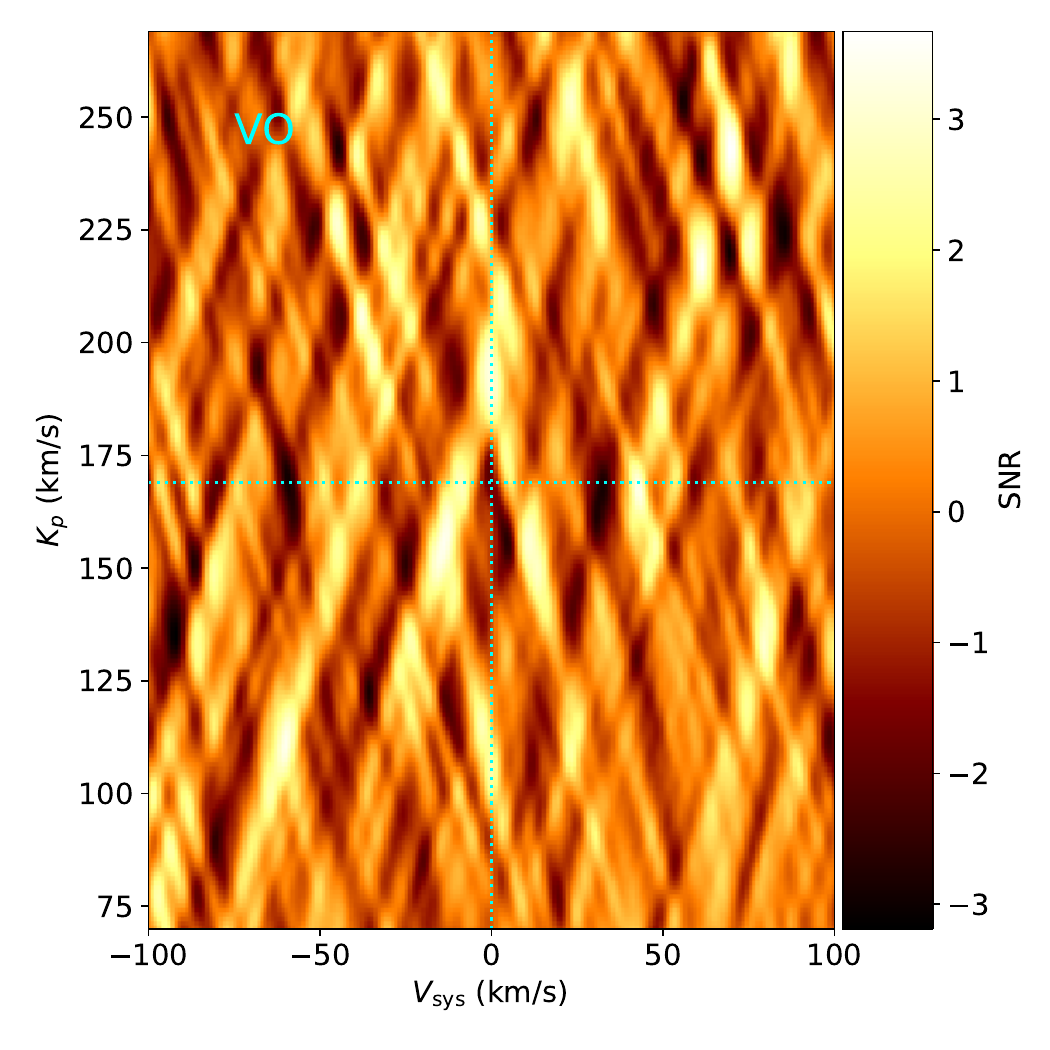}\includegraphics[width=0.4\textwidth]{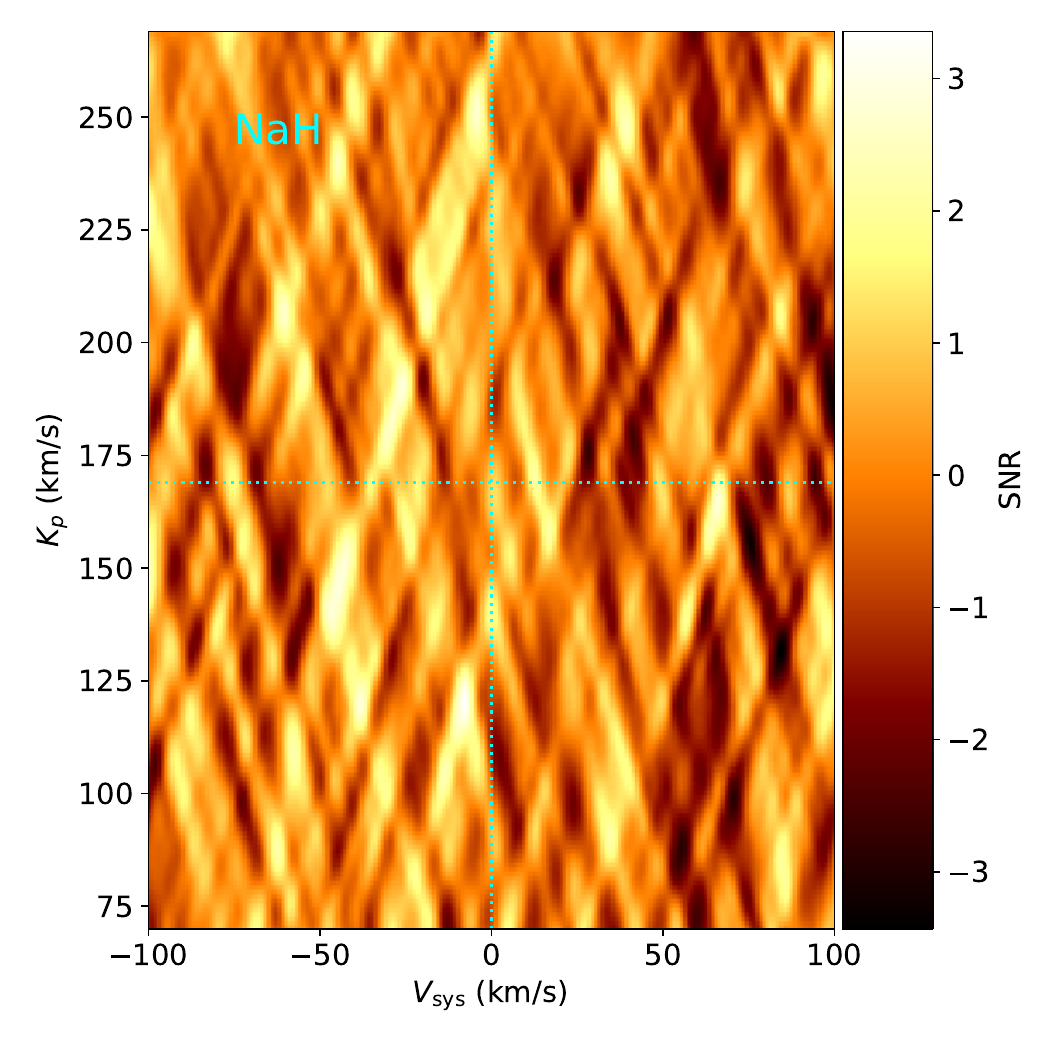} \\ 
	\includegraphics[width=0.4\textwidth]{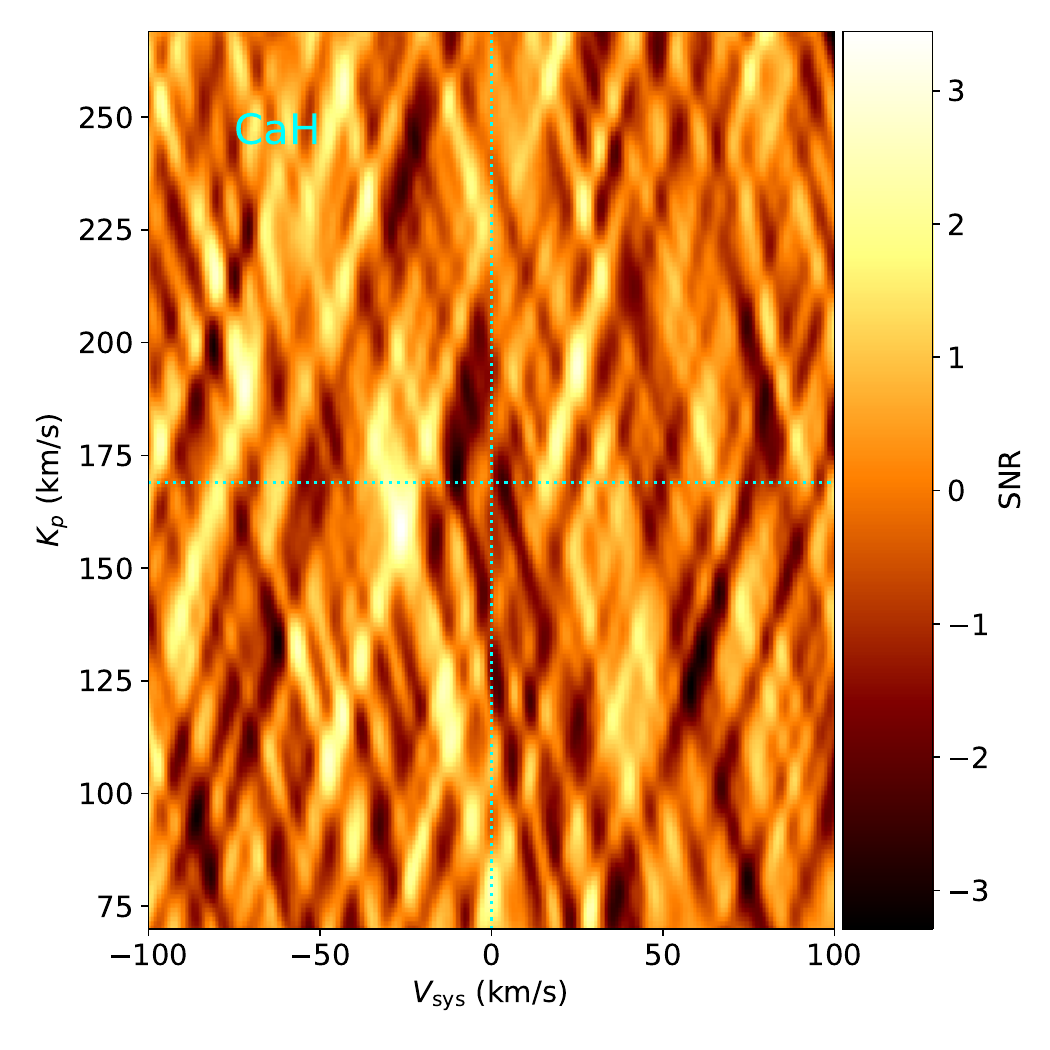}\includegraphics[width=0.4\textwidth]{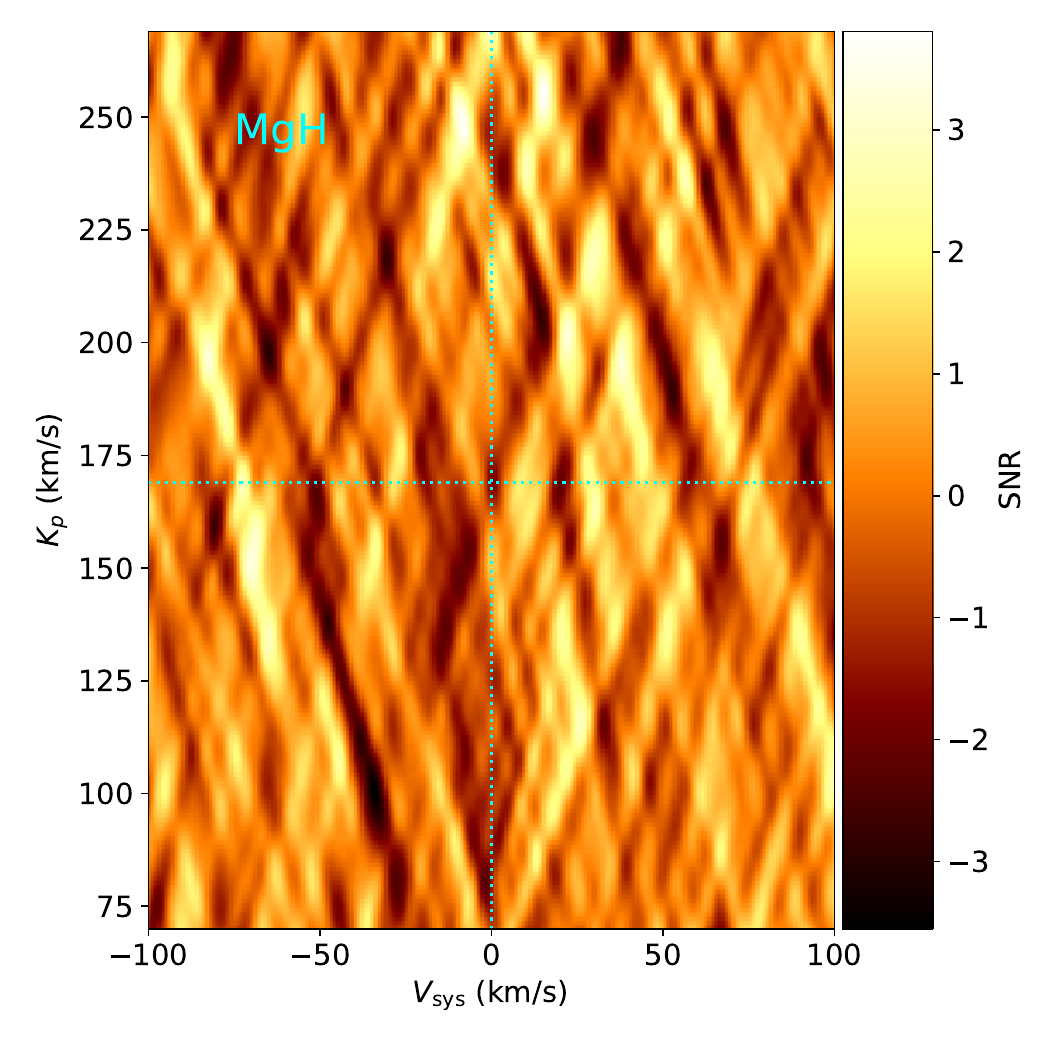} \\ 
	\includegraphics[width=0.4\textwidth]{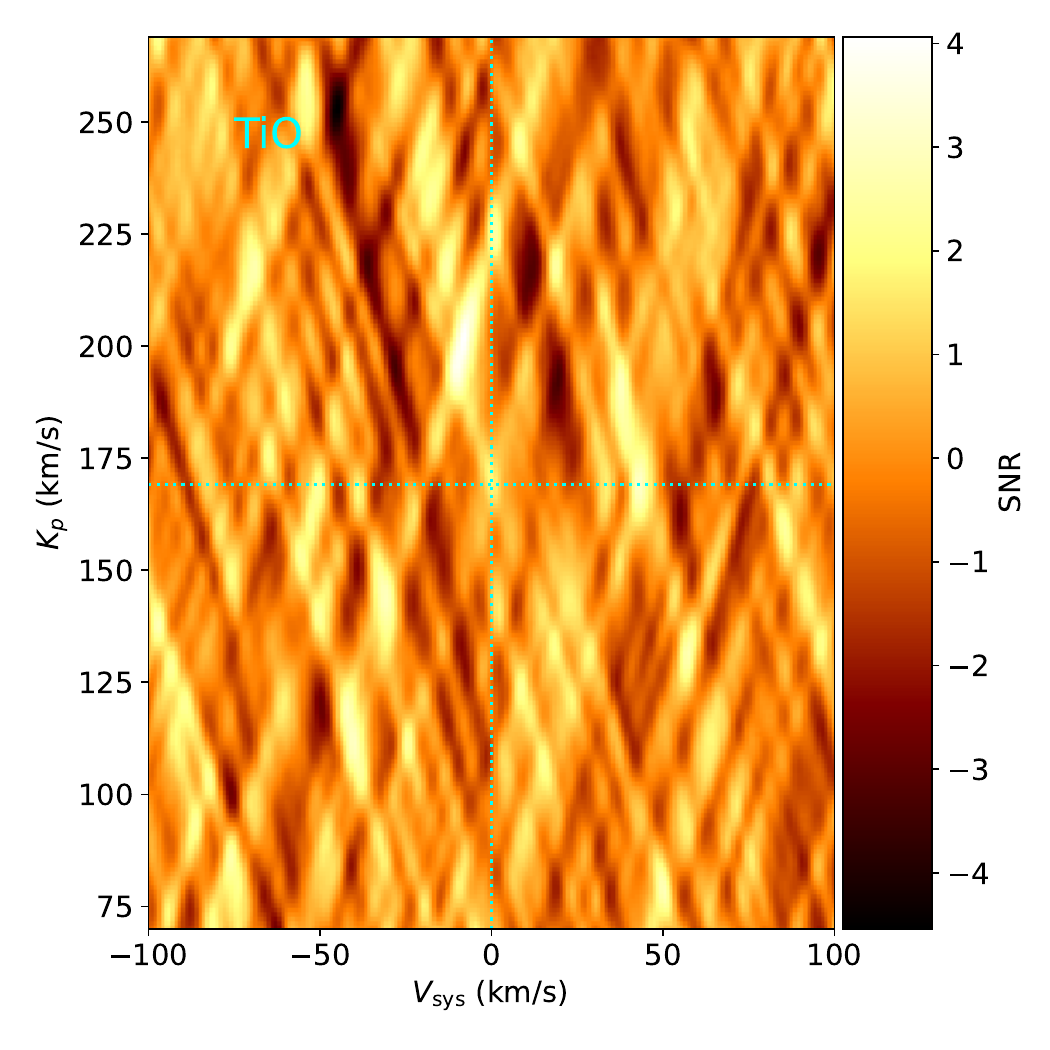}
    \caption{Shifted and combined CCF maps for VO, NaH, CaH, MgH, and TiO. These species yielded high detectability values with our atmospheric models, but did not result in a detection.}
    \label{fig:additionalnon-detections}
\end{figure*}

We show the CCF maps for several molecular species with high detectability values in Fig.~\ref{fig:additionalnon-detections}. All resulted in non-detections; VO, CaH, and TiO confirm the non-detections from \cite{Johnson2022}, while NaH and MgH are presented here for the first time.

\bsp	
\label{lastpage}
\end{document}